\documentclass[useAMS,usenatbib]{mn2e}
\usepackage{natbib}
\usepackage{supertabular}
\usepackage{pdflscape}
\usepackage{graphicx}
\usepackage{amsmath}
\usepackage{color}

\def\ca2T{Ca\,{\sc{ii}} Triplet}

\def\msun{M$_\odot$}

\def\aj{AJ}%
\def\actaa{Acta Astron.}%
\def\apj{ApJ}%
\def\apjl{ApJ}%
\def\apjs{ApJS}%
\def\aap{A\&A}%
\def\mnras{MNRAS}%
\title[Signatures of tidally stripped stellar populations from the inner Small Magellanic Cloud]
{The VMC Survey - XXIV. Signatures of tidally stripped stellar populations from the inner Small Magellanic Cloud \thanks{Based on observations made with VISTA at ESO under programme ID 179.B-2003.}}
\author[Subramanian et al.]
{Smitha Subramanian,$^1$\thanks{E-mail:
smithaharisharma@gmail.com} Stefano Rubele,$^{2,3}$ Ning-Chen Sun,$^{1,4}$ L\'eo Girardi,$^{2}$ 
\newauthor
Richard de Grijs,$^{1,4,5}$ Jacco Th. van Loon,$^{6}$  Maria-Rosa L. Cioni, $^{7,8,9}$ 
\newauthor
Andr\'es E. Piatti,$^{10,11}$ Kenji Bekki,$^{12}$ Jim Emerson,$^{13}$ Valentin D. Ivanov,$^{14,15}$ 
 \newauthor
Leandro Kerber,$^{16}$ Marcella Marconi,$^{17}$ Vincenzo Ripepi $^{17}$ \& Benjamin L. Tatton$^{6}$\\\\
$^{1}$Kavli Institute for Astronomy and Astrophysics, Peking University, Yi He Yuan Lu 5, Hai Dian District, Beijing 100871, China\\
$^{2}$Osservatorio Astronomico di Padova -- INAF, Vicolo dell'Osservatorio 5, I-35122 Padova, Italy\\
$^3$Dipartimento di Fisica e Astronomia, Universit\`a di Padova, vicolo dell'Osservatorio 2, Padova I-35122, Italy\\
$^{4}$Department of Astronomy, Peking University, Yi He Yuan Lu 5, Hai Dian District, Beijing 100871, China\\
$^{5}$International Space Science Institute -- Beijing, 1 Nanertiao, Zhongguancun, Hai Dian District, Beijing 100190, China\\
$^{6}$Lennard-Jones Laboratories, Keele University, Staffordshire, ST5 5BG, UK\\
$^{7}$Institut f\"{u}r Physik und Astronomie, Universit\"{a}t Potsdam, Karl-Liebknecht-Str. 24/25, D-14476 Potsdam, Germany\\
$^{8}$Leibnitz-Institut f\"{u}r Astrophysik Potsdam, An der Sternwarte 16, D-14482 Potsdam, Germany\\
$^{9}$University of Hertfordshire, Physics Astronomy and Mathematics, College Lane, Hatfield AL10 9AB, UK\\
$^{10}$Observatorio Astron\'omico, Universidad Nacional de C\'ordoba, Laprida 854, 5000, C\'ordoba, Argentina\\
$^{11}$Consejo Nacional de Investigaciones Cient\'{\i}ficas y T\'ecnicas, Av. Rivadavia 1917, C1033AAJ, Buenos Aires, Argentina\\
$^{12}$ICRAR, M468, The University of Western Australia, 35 Stirling Hwy, Crawley, WA 6009, Australia\\
$^{13}$Astronomy Unit, School of Physics and Astronomy, Queen Mary University of London, Mile End Road, London E1 4NS, UK\\
$^{14}$European Southern Observatory, Ave. Alonso de C\'ordova 3107, Vitacura, Santiago, Chile \\
$^{15}$European Southern Observatory, Karl-Schwarzschild-Str. 2, D-85748 Garching bei M\"{u}nchen, Germany\\
$^{16}$Universidade Estadual de Santa Cruz, Rodovia Ilh\'eus-Itabuna, km 16, 45662-200 Ilh\'eus, Bahia, Brazil\\
$^{17}$INAF-Osservatorio Astronomico di Capodimonte, Via Moiariello 16, I-80131 Naples, Italy\\}

\begin{document}
 \date{Received/Accepted}

\pagerange{\pageref{firstpage}--\pageref{lastpage}} \pubyear{...}

\maketitle

\label{firstpage}
\begin{abstract}

We study the luminosity function of intermediate-age red-clump stars using deep, near-infrared photometric data covering $\sim$ 20 deg$^2$ located throughout the central part of the Small Magellanic Cloud (SMC), comprising the main body and the galaxy's eastern wing, based on observations obtained with the VISTA Survey of the Magellanic Clouds (VMC). We identified regions which show a foreground population ($\sim$11.8 $\pm$ 2.0 kpc in front of the main body) in the form of a distance bimodality in the red-clump distribution. The most likely explanation for the origin of this feature is tidal stripping from the SMC rather than the extended stellar haloes of the Magellanic Clouds and/or tidally stripped stars from the Large Magellanic Cloud. The homogeneous and continuous VMC data trace this feature in the direction of the Magellanic Bridge and, particularly, identify (for the first time) the inner region ($\sim$ 2 -- 2.5 kpc from the centre) from where the signatures of interactions start becoming evident. This result provides observational evidence of the formation of the Magellanic Bridge  from tidally stripped material from the SMC.  
\end{abstract}

\begin{keywords}
stars: individual: red clump stars - Magellanic Clouds - galaxies: interactions.
\end{keywords}

\section{Introduction}
The Magellanic system, which comprises of the Large Magellanic Cloud (LMC), the Small Magellanic Cloud (SMC), the connecting stream of gas and stars known as the Magellanic Bridge (MB), the leading stream of neutral hydrogen known as the leading arms and the trailing stream of gas, the Magellanic Stream (MS), is one of the nearest examples of an interacting system of galaxies in the local Universe. The LMC and SMC, together known as the Magellanic Clouds (MCs), are two nearby galaxies located at a distance of 50$\pm$1 kpc (e.g. \citealt{degrijs2014}; \citealt{Elgueta2016}) and 61$\pm$1 kpc (e.g. \citealt{degrijs2015}) respectively.  
The SMC is located $\sim$ 20$^\circ$ west of the LMC on the sky. The total mass of the LMC and the SMC estimated from the rotation curves is 1.7 $\times$ 10$^{10}$ \msun (up to a radius of 8.7 kpc; \citealt{vanderMarel2014}; stellar mass of $\sim$ 2.7 $\times$ 10$^9$ M$_\odot$ and gas mass of $\sim$ 5 $\times$ 10$^8$ \msun) and  2.4 $\times$ 10$^9$ \msun (\citealt{SS2004}; stellar mass of $\sim$ 4.2 $\times$ 10$^8$ M$_\odot$ and gas mass of $\sim$ 3.1 $\times$ 10$^8$ \msun), respectively. Based on these mass estimates, the baryonic fraction of the MCs (LMC: $\sim$ 18\% and SMC: $\sim$ 30\%) is  larger than the typical value (3 -- 5 \%) observed for Milky Way type galaxies. Hence, \cite{Besla2015} suggested that the total masses of the MCs should be $\sim$ 10 times larger than those which are traditionally estimated. 

The MCs are known to have had interactions with each other as well as with the Milky Way (MW).  The MB, MS and the leading arm, which are prominent in H\,{\sc i} maps \citep{Putman2003}, are signatures of these interactions. It is also believed that the tidal forces caused by these interactions have caused structural changes in the MCs as well as in the Galaxy.  
The MCs host a range of stellar populations, from young to old. Their proximity allows us to resolve individual stars and their location well below the Galactic plane makes them less affected by Galactic reddening, enabling us to probe faint populations. Those stellar populations which are standard candles, like Cepheids, RR Lyrae stars and red-clump (RC) stars, can be used to study the 3D structure of the system and to identify regions which show signatures of interaction as structural deviations. Thus, the Magellanic system is an excellent template to study galaxy interactions using resolved stellar populations. 

Recent models (\citealt{Besla2010,Besla2012,Besla2013}; {\citealt{DB2012}) based on new proper motion estimates of the MCs explain many of the observed features of the MS and leading arm as having originated from the mutual tidal interaction of the MCs.  \cite{Hammer2015} proposed a model based on ram-pressure forces exerted by the Galactic corona and collision of the MCs to explain their origin.  The observed properties of the MS and leading arm are not fully explained by a single model. Hence, their origin and especially the role of the Milky Way in the formation of the Magellanic system is not fully understood. A detailed review of the formation of the MS and the interaction history of the MCs is given by  \cite{FOX2015}. 

On the other hand, the origin of the MB is fairly well explained by these models as the result of the tidal interaction of the MCs during their recent encounter $\sim$ 100 -- 300 Myr ago. 
The MB contains very young stellar populations (\citealt{DB1998}; {\citealt{Harris2007}; \citealt{Chen2014}; \citealt{skowron2014}, and references therein), which might have formed from the gas stripped during the interaction. Tidal forces have similar effects on both the stars and the gas in a system and, hence, an interaction between the MCs $\sim$ 100 -- 300 Myr ago should have affected stars in the MCs older than 300 Myr. Thus, an intermediate-age/older stellar population which would have been stripped during the interaction is expected to be present in the MB. Earlier studies by \cite{DB1998} and \cite{Harris2007} did not find the presence of intermediate-age/old stellar populations in fields centred on the H\,{\sc i} ridge line of the MB. However, recent studies (\citealt{BCN2013}; \citealt{Noel2013, Noel2015}; \citealt{skowron2014}) found evidence of the presence of intermediate-age/old stellar populations in the central and western regions of the MB.  

Earlier spectroscopic studies of young stars in the MB suggested that their progenitor material contains contributions from the SMC, both of enriched and less enriched gas (\citealt{Hambly1994}; \citealt{Rolleston1999}; \citealt{Dufton2008}). Investigating the RC stars in the outer regions ($>$ 2$^\circ$ from the centre) of the SMC, \cite{Hatzidimitriou1989} found large line-of-sight depths in the north-eastern regions and a follow-up spectroscopy (\citealt{HCH1993}), identified a correlation between distance and radial velocity. The authors interpreted this as the effect of tidal interactions between the MCs.
\cite{Piatti2015} analysed the star clusters in the outermost eastern regions of the SMC and found an excess of young clusters, which could have been formed during recent interactions of the MCs and \cite{Bica2015} found that many of these clusters are closer to us than the SMC's main body.  

\cite{Nid2013} identified a closer (distance, d $\sim$ 55 kpc) stellar structure in front of the main body of the eastern SMC, which is located 4$^\circ$ from the SMC centre. These authors suggested that it is the tidally stripped stellar counterpart of the H\,{\sc i} in the MB. From their analysis of a few fields in the MB,  \citet{Noel2013, Noel2015} found from the synthetic colour--magnitude diagram (CMD) fitting technique that the intermediate-age population in the MB has similar properties to stars in the inner $\sim$ 2.5 kpc  region of the SMC and suggested that they were tidally stripped from the inner SMC region. Meanwhile,  \cite{skowron2014}, who studied a more complete region of the MB, explained the presence of intermediate-age stars as the overlapping haloes of the MCs. A spectroscopic study of red giant branch (RGB) stars by \cite{Olsen2011} reported a population of tidally accreted SMC stars in the outer regions of the LMC. The inner 2$^\circ$ region of the SMC did not show any evidence of tidal interactions \citep{HZ2006}, whereas the outer regions show evidence of substantial stripping owing to interactions \citep{Dobbie2014}. \cite{Dobbie2014} found kinematic evidence (stars with lower line-of-sight velocities) of tidally
stripped stars associated with the MB. Thus, most of these studies support tidal stripping of stars/gas from the SMC. 

\cite{Depropris2010} also found the velocity distribution of the RGB stars to the East and South of the SMC centre to be bimodal. However, contrary to \cite{Dobbie2014}, they found the second peak at a larger line-of-velocity than that of the mean component, indicating that the stellar populations in the eastern SMC have properties similar to those of the LMC. Their fields also show a large range of metallicities, even similar to that of the LMC. This suggests that the LMC stars may also be tidally affected by the interaction and were accreted to the SMC. Thus, the nature and origin of the intermediate-age/old populations in the MB is not well understood. 

Most studies which have tried to understand the interaction history of the MCs and the formation of the MB concentrated on the outer regions of these galaxies. Studies using homogeneous and continuous data from the inner to the outer regions of the SMC (where the stellar density is high) are essential to understand the effects of tidal interactions on stars. 
In this context, we present a detailed analysis of RC stars in the inner regions for $\it{r}$ $\le$ 4$^\circ$ ($\sim$ 20 deg$^2$) of the SMC using photometric data from the VISTA (Visible and Infrared Survey Telescope for Astronomy) survey of the Magellanic Clouds (VMC; \citealt{Cioni2011}).  The RC stars are metal-rich counterparts of the horizontal-branch stars. They have an age range of 2 -- 9 Gyr and a mass range of 1 -- 3 \msun (\citealt{Girardi2001}; \citealt{Girardi2016}). They have a relatively constant colours and magnitudes, which make them useful to study the 3D structure and reddening (e.g.\citealt{Tatton2013}) of their host galaxies.  Along with the inner 2$^\circ$ region, which was covered by previous optical surveys, the VMC data studied here partially cover the 2$^\circ$ -- 4$^\circ$ region of the SMC which is not well explored. Thus, the homogeneous and continuous VMC data used in the present study are unique tools to explore the effect of tidal interactions and the presence of tidally stripped stars in the inner SMC. 
\begin{figure}
\includegraphics[width=0.52\textwidth,height=0.4\textwidth]{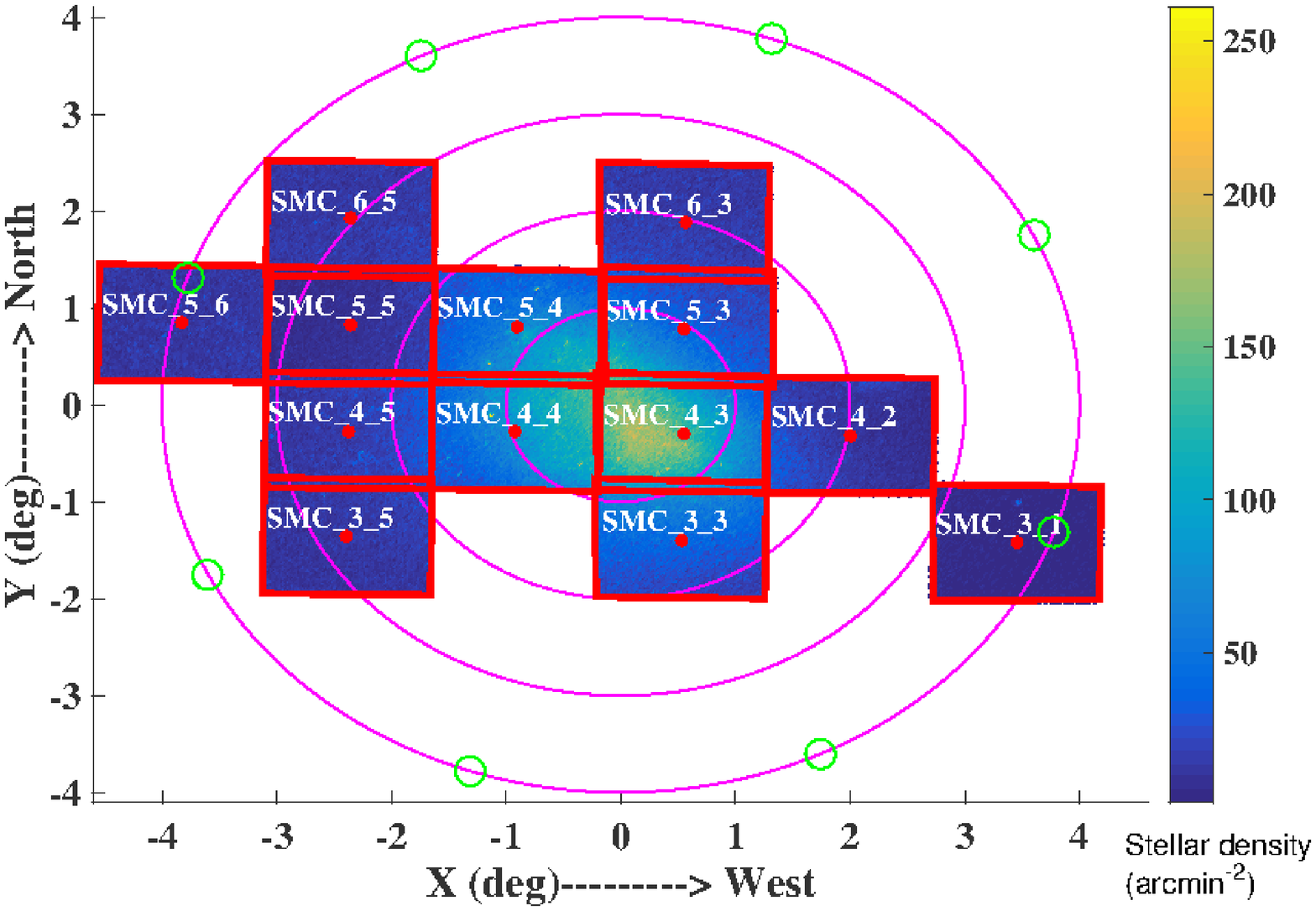}
\caption{The positions of the VMC tiles analysed in this study are shown in the $XY$ plane. $X$ and $Y$ are defined as in \protect\cite{VC2001} with centre co-ordinates at $\alpha_0$ = 00$^h$52$^m$12$^s$.5 and $\delta_0$ = $-$72$^{\circ}$49$^\prime$43$^{\prime\prime}$(J2000; \citealt{DV1972}). The concentric circles represent 1$^\circ$, 2$^\circ$, 3$^\circ$ and 4$^\circ$ radii the centre of the SMC. The colour code from blue to yellow represents increasing stellar density. The central co-ordinates of the locations of the 8 fields \citep{Nid2013} at a radius of 4$^\circ$ from the SMC centre are shown as green open circles.}
\vspace{-0.5cm}
\end{figure} 

\section{VMC Data}
The VMC survey is a continuous and homogeneous ongoing survey of the Magellanic system in the $\it{YJK_\textnormal{s}}$ (central wavelengths, $\lambda_c$ = 1.02 $\mu$m, 1.25 $\mu$m and 2.15 $\mu$m, respectively) near-infrared (NIR) bands using the 4.1 m VISTA telescope located at Paranal Observatory in Chile. On  completion, the survey is expected to cover $\sim$ 170 deg$^2$ (LMC: 105 deg$^2$, SMC: 42 deg$^2$, MB: 20 deg$^2$ and MS: 3 deg$^2$) of the Magellanic system. The limiting magnitudes with signal-to-noise (S/N) ratio of 5, for single-epoch observations of each tile, in the $\it{Y, J} $ and $\it{K_\textnormal{s}}$ bands are $\sim$ 21.1 mag, 20.5 mag and 19.2 mag, respectively, in the Vega system. The stacked images can provide sources with limiting magnitudes of up to 21.5 mag in $\it{K_\textnormal{s}}$ with S/N = 5. 
The RC feature in the SMC ($\it{K_\textnormal{s}}$ $\sim$ 17.3 mag) is around 2 mag brighter than the 5 $\sigma$ detection limit of single-epoch observations. 

In the present work, we studied the 13 tiles (each covers an area of $\sim$ 1.6 deg$^2$) of the SMC, which cover $\sim$ 20 deg$^2$ and comprise both the main body of the SMC and the eastern wing. Fig. 1 shows the stellar density distribution of these tiles in the $\it{XY}$ plane. $\it{X}$ and $\it{Y}$ are defined as in \cite{VC2001} with centre coordinates at 
$\alpha_0$ = 00h 52m12s .5 and $\delta_0$ = $-$72$^{\circ}$49$^\prime$43$^{\prime\prime}$(J2000; \citealt{DV1972}). The concentric circles represent radii of 1$^\circ$, 2$^\circ$, 3$^\circ$ and 4$^\circ$ from the centre of the SMC. 
The data used were retrieved from the VISTA Science Archive (VSA; \citealt{Cross2012}). 
Point-spread-function (PSF) photometry of 10 tiles (except tiles SMC\_5\_5, 4\_2 and 3\_1) was performed by \cite{Rubele2015} and we used these PSF photometric data in the $\it{Y}$ and $\it{K_\textnormal{s}}$ bands for our analysis. For the remaining three tiles, PSF photometry was performed following the steps explained in the appendix of \cite{Rubele2015}. From the PSF catalogues, we selected the most probable stellar sources based on the sharpness criteria ($-1$ to 1 because those below $-1$ are likely bad pixels and those above 1 are likely extended sources) and photometric errors ($\le$ 0.15 mag). The RC feature in the SMC is $\sim$ 3 to 3.5 mag brighter than the 50\% photometric completeness limit and the typical photometric uncertainty associated with RC magnitudes is $\sim$ 0.05 mag in the $\it{K_\textnormal{s}}$ band. 

\section{Selection of RC stars}
We use the $\it{Y}$ and $\it{K}_{s}$ band photometric data to do the selection and further analysis of the RC stars. The $\it{(Y-K_\textnormal{s})}$ colour gives the widest colour separation possible in the VMC data and hence allows a better separation between RC and RGB stars. The $\it{K_\textnormal{s}}$-band magnitudes of the RC stars are less affected by extinction and population effects \citep{SG2002}. Hess diagrams, with bin sizes of 0.01 mag in $\it{(Y-K_\textnormal{s})}$ colour and 0.04 mag in $\it{K_\textnormal{s}}$, representing the stellar density in the observed $\it{(Y-K_\textnormal{s})}$ vs $\it{K_\textnormal{s}}$ CMD of the 13 tiles are shown in Fig. 2. The blue to red colour code represents the increasing stellar density in each tile. The RC stars, at $(Y-K_\textnormal{s})$ $\sim$ 0.7 mag and $\it{K_\textnormal{s}}$ $\sim$ 17.3 mag, are easily identifiable in all panels.

We initially defined a box with size, 0.5 $\le$ $(Y-K_\textnormal{s})$ $\le$ 1.1 mag in colour and 16.0 $\le$ $\it{K_\textnormal{s}}$ $\le$ 18.5 mag in magnitude, to select the RC stars from the CMD. The selection box is shown in all panels of Fig. 2. The range in magnitude was chosen to include the vertical extent (clearly seen in tiles SMC\_5\_6 and SMC\_6\_5) of the RC regions and also to incorporate the shift towards fainter magnitudes due to interstellar  extinction.  
\begin{landscape}
\begin{figure}
\centering
\includegraphics[height=0.99\textwidth,width=1.55\textwidth]{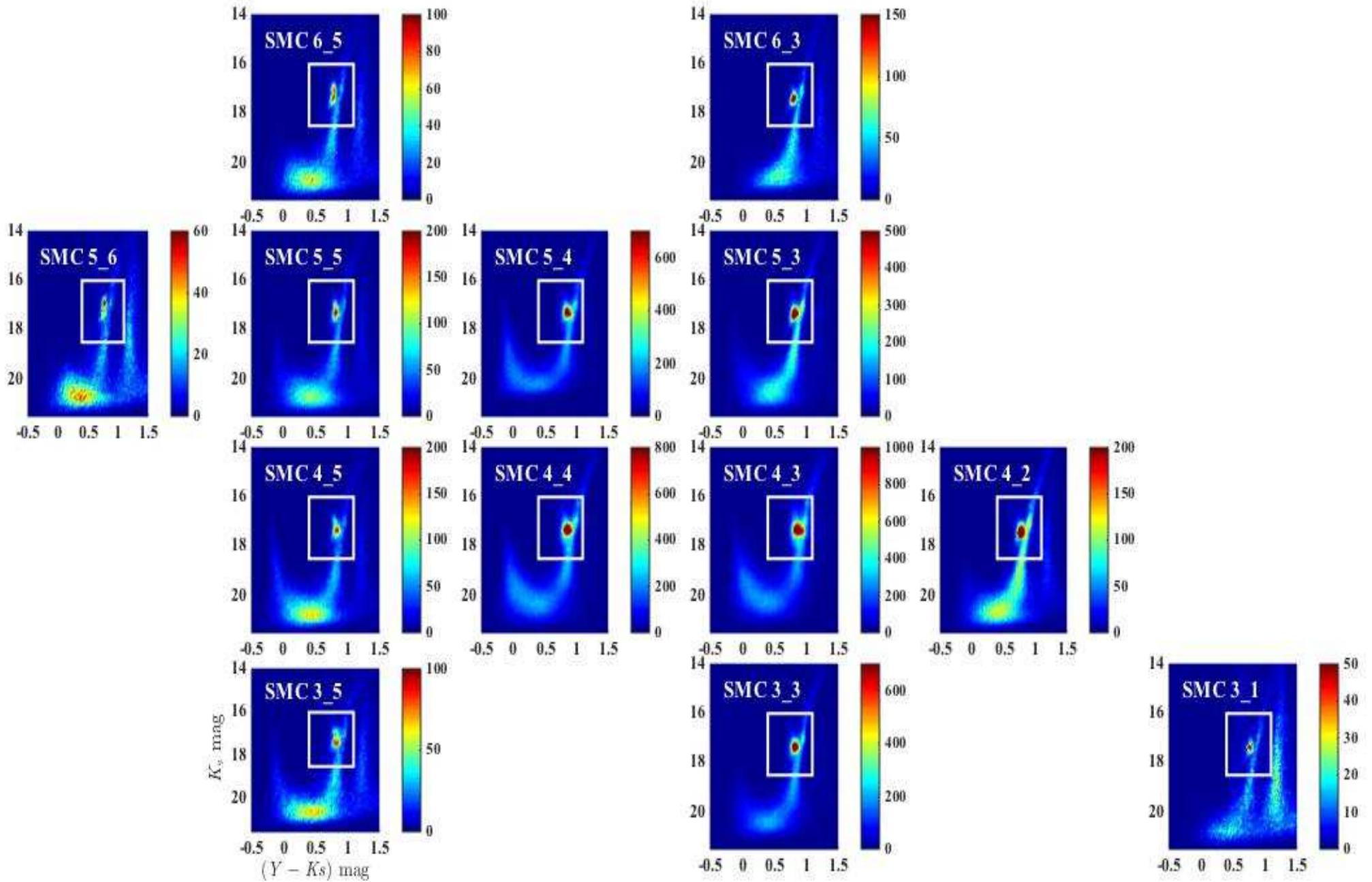}
\caption{Hess diagrams representing the stellar density in the observed CMD of the SMC tiles. The blue to red colour code represents the increasing stellar density in each tile. The white box represents the initial selection box of the RC stars. }
\end{figure}
\end{landscape}
As the colour cut which separates the RC and RGB is not very well defined from the CMDs, especially in the crowded central regions, RGB stars are also included in the initial selection box. Below, we will model the RGB density and subtract it from the Hess diagrams to obtain the RC distribution (see Sect. 4). The following steps describe the MW foreground removal and the reddening/extinction correction of stars in the box.\\

1) {\bf Removal of Galactic foreground stars:} The $\it{Y}$ and $K_\textnormal{s}$ magnitudes of Galactic foreground stars towards the SMC tiles are obtained using the TRILEGAL MW stellar population model \citep{Leo2005} which includes a model of extinction in the Galaxy.  A typical Galactic foreground contribution in the $(Y-K_\textnormal{s})$ vs $\it{K_\textnormal{s}}$ CMD towards the SMC region is shown in Fig. 3. The contribution of MW stars in the RC box region of the CMD has a range from 1\% to 30\% of the total number of stars in the box, depending on the location within the SMC.  The RC region in the CMD of each tile is cleaned by removing the closest matching star corresponding to the colour and magnitude of each MW star from the model. \\

2) {\bf Extinction correction:} The cleaned CMD region is then corrected for foreground and internal extinction using the extinction map of \cite{Rubele2015}. They only report extinction values for 10 of the 13 tiles (SMC 3\_1, SMC 4\_2 and SMC 5\_5 are not included). In their study, each tile is divided into 12 sub-regions and a synthetic CMD technique is applied to retrieve the star-formation history, metallicities, distances and extinction values of each sub-region.  The extinction (A$_{\it{K_\textnormal{s}}}$) values have a range from 0.04 -- 0.07 mag. The average extinction they obtained for the analysed SMC region is $\it{A_{Ks}}$ = 0.05 $\pm$ 0.01 mag. The variation of extinction within a tile is $\sim$ $\it{A_{Ks}}$ = 0.01 mag. The parameters these authors derived are mainly sensitive to the intermediate-age population in the SMC.  We applied the extinction values corresponding to the sub-region in which the RC stars fall and corrected for their extinction. For the stars in the remaining three tiles we applied the extinction values of the nearest sub-region. 

We note that the Galactic foreground is very smooth in the RC region of the CMD, and that the extinction correction is in general much smaller than the extent of the RC features discussed later in this paper. It is therefore unlikely that these steps could introduce any significant error or bias in our subsequent analysis.

\begin{figure}
\hspace{-0.5cm}
\includegraphics[width=0.55\textwidth,height=0.65\textwidth]{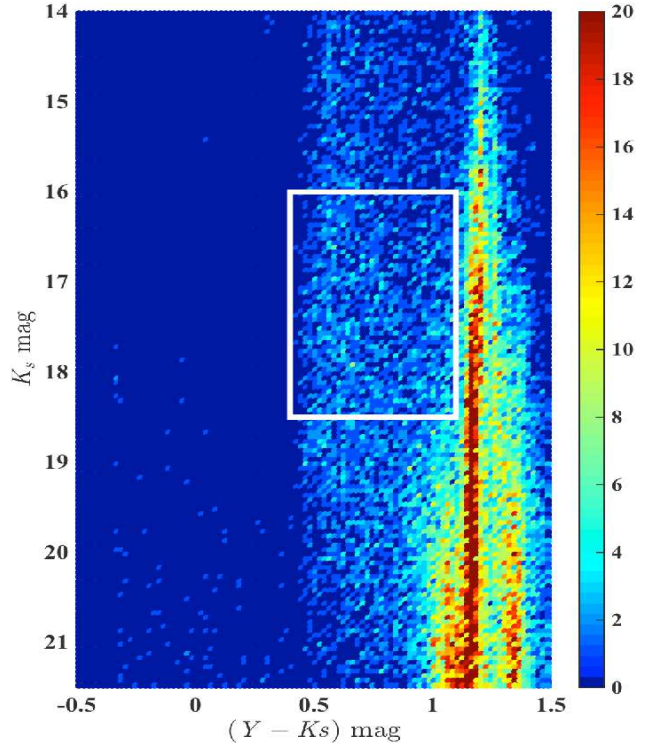}
\vspace{-0.9cm}
\caption{Hess diagram representing the stellar density in the observed CMD of the typical MW foreground stars towards the SMC. The RC selection box from Fig. 1 is also shown.}
\end{figure}

\section{RC luminosity function and bimodality} 
Figure 4 shows the RC region of the Hess diagrams after removing the MW foreground and correcting for reddening/extinction. These are normalised Hess diagrams with respect to the bin with the maximum number of stars, and the colour code, from blue to red, represents the increase in the stellar density in each region. This figure  clearly shows the vertical extent/double clump feature in the eastern tiles (SMC 5\_6, SMC 6\_5, SMC 5\_5, SMC 4\_5 and SMC 3\_5). To understand this feature better, we divided each tile into four equal sized
 ($\sim$ 0.75 $\times$ 0.55 deg$^2$) sub-regions. The luminosity functions of the RC stars in each sub-region are analysed separately.
 
\begin{figure*}
\includegraphics[width=0.7\textwidth,height=1.0\textwidth,angle=270]{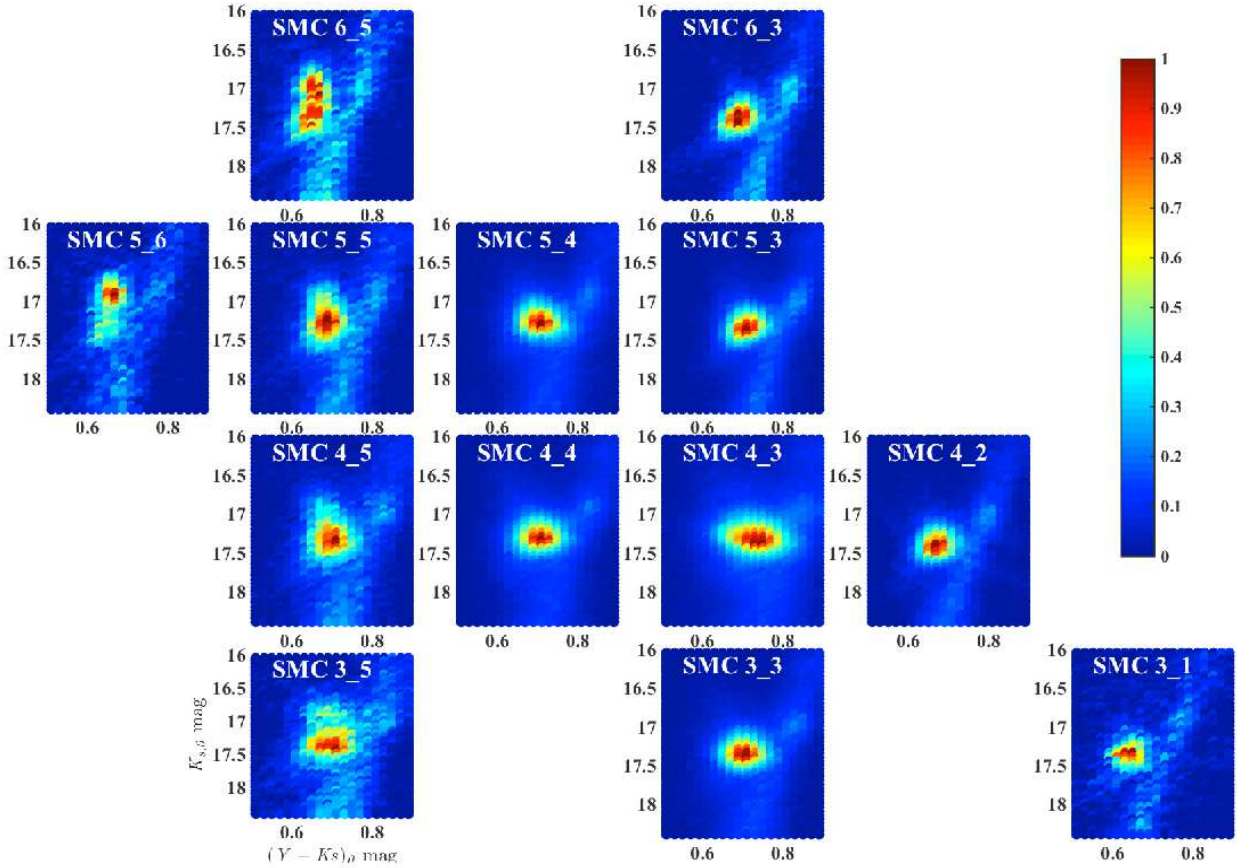}
\caption{ Reddening-corrected and MW-foreground-subtracted RC region in the $(Y-K_\textnormal{s})_0$ vs $K_\textnormal{s,0}$ CMD of the SMC fields. The colour bar shows the normalised stellar density with respect to the bin with the maximum number of stars in each region. The axis labels are only shown for the bottom left panel.}
\end{figure*}
 
 We performed a careful analysis to remove RGB contamination in order to analyse the RC luminosity function. The RC stars in each sub-region were identified and Hess diagrams similar to Fig. 4 were constructed. The bin sizes in colour and magnitude were 0.01 mag and 0.04 mag, respectively. The separation of the RC from the RGB  based on their colour may not be reliable for the fainter magnitudes and we restrict the analysis of the luminosity function up to $K_\textnormal{s,0}$ = 18 mag. From Fig. 4 we can see that this fainter magnitude cut off of  $K_\textnormal{s,0}$ = 18 mag is $\sim$ 0.6 -- 0.7 mag lower than the peak of the main RC. In a given magnitude bin (along a row in the Hess diagram), we performed a double Gaussian profile fit to the colour distribution and obtained the peak colours corresponding to the RC and the RGB distributions. This was repeated for all magnitude bins, over the range of 16.0 -- 18.0 mag in the $K_\textnormal{s,0}$ band.  After this first step, we performed a linear least-squares fit to the RGB colours with the magnitude corresponding to each bin. Again, the first step was repeated with the constraint that the colour of the RGB corresponding to each magnitude is the same as that obtained from the linear fit. From the resultant Gaussian parameters corresponding to the RGB, we obtained a model for the RGB density distribution. This model was subtracted from the Hess diagram and a clean RC distribution is obtained. 

The RGB model and the RGB-subtracted RC distribution for the four sub-regions of the tile SMC 6\_5 are shown in Fig. 5. The locations of these sub-fields are such that the sub-regions, (SMC 6\_5\_a and SMC 6\_5\_b) and (SMC 6\_5\_c and SMC 6\_5\_d) are in the western and eastern regions of tile SMC 6\_5, respectively. A similar procedure was applied to all tiles in our study. For tile SMC 3\_1, the number of stars in each sub-region was not sufficient to perform reasonable fits to model the RGB. Therefore, we divided the tile into 2 sub-regions instead of 4.  The final RC density distribution is summed in the colour range 0.55 $\le$ $(Y-K_\textnormal{s})_{0}$ $\le$ 0.85 mag to obtain the luminosity function of the RC stars. 

Initially, the luminosity function was modelled with a single Gaussian profile to account for the RC stars and a quadratic polynomial term to account for the remaining RGB contamination. The profile parameters, associated errors and the reduced $\chi^2$ were obtained. An additional Gaussian component to account for the secondary peak in the RC distribution was included and we performed the fit again to obtain the profile parameters, the associated errors and the reduced $\chi^2$ value. If the reduced $\chi^2$ improved by more than 25\% after including the second Gaussian component, then it was considered a real  component. This choice of reduced $\chi^2$ naturally removes any second Gaussian component with peak less than twice the mean residual of the total fit. Most of the sub-regions showed an improvement of $\sim$ 25\% -- 50\% in the reduced $\chi^2$ values when an additional Gaussian component was included. The RC luminosity function and the Gaussian profile fits of the four sub-regions of the field SMC 6\_5 are illustrated in Fig. 5. We can see that all sub-fields in tile SMC 6\_5 show bimodality in the luminosity function. 

The final luminosity functions of the RC stars in all 50 sub-fields are shown in Fig. 6. The Gaussian parameters of the profile fits are given in Table 1. From Table 1 and Fig. 6, we can see that all tiles have a Gaussian component corresponding to a $K_\textnormal{s,0}$ value of $\sim$ 17.3 -- 17.4 mag. The prominent and interesting feature in Fig. 6 is that all eastern tiles with tile centres $\it{r}$ $\ge$ 2$^{\circ}$.5 (SMC 5\_6, SMC 6\_5, SMC 5\_5, SMC 4\_5 and SMC 3\_5) show bimodality in the luminosity function, as indicated by the double Gaussian profiles (two peaks at $\sim$ 17.3 -- 17.4 mag and at $\sim$ 16.9 -- 17 mag).  Tiles SMC 4\_3, SMC 4\_4, SMC 5\_3, SMC 5\_4, SMC 3\_3, SMC 6\_3 and two sub-regions of SMC 4\_2 show a broad component along with the narrow component.  The peaks of the broad components are slightly brighter ($\sim$ 0.05 -- 0.15 mag) than the peak of the narrow component. Tile SMC 3\_1 and two western sub-regions of SMC 4\_2 show only a single Gaussian component. 
The nature and cause of the observed brighter component in the eastern tiles are discussed in detail in the next section.

\begin{figure*}
\centering
\setlength{\tabcolsep}{0.000001mm}
\renewcommand{\arraystretch}{0.00001}
\begin{tabular}{ll}
\includegraphics[height=0.55\textwidth,width=0.52\textwidth]{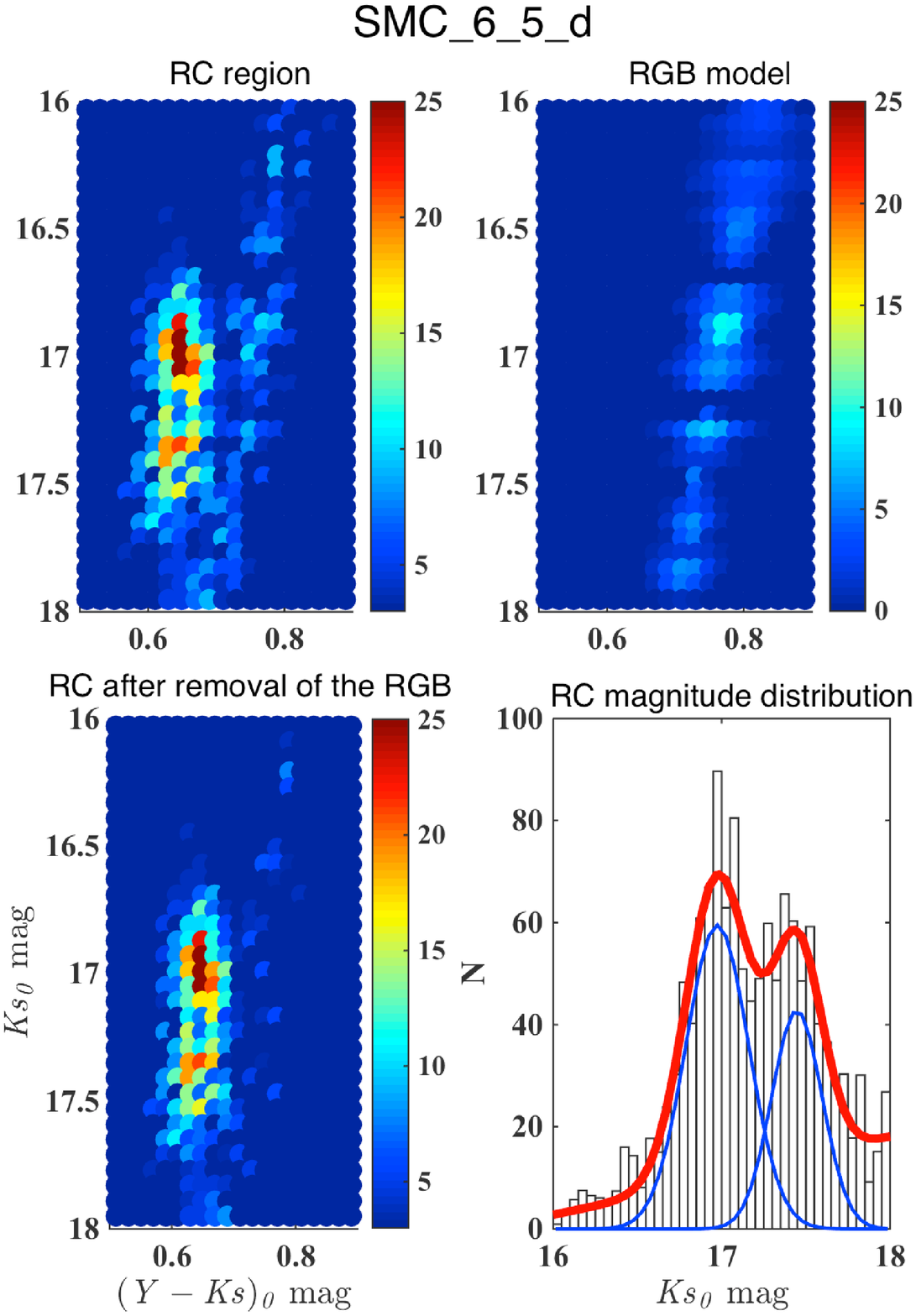} 
 &
 \hspace{-0.85cm}
\includegraphics[height=0.55\textwidth,width=0.52\textwidth]{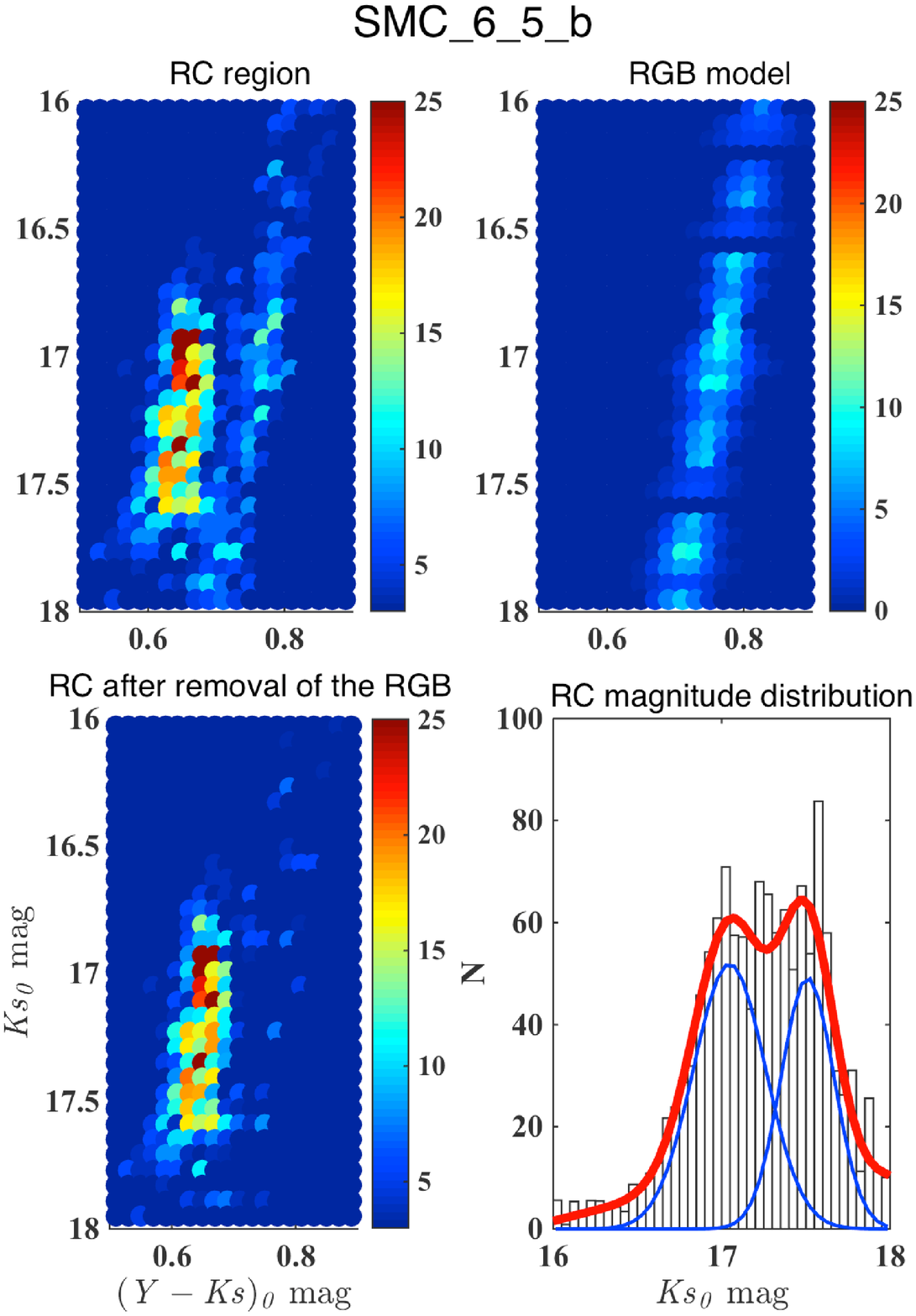} \vspace{-0.4cm}\\

\includegraphics[height=0.55\textwidth,width=0.52\textwidth]{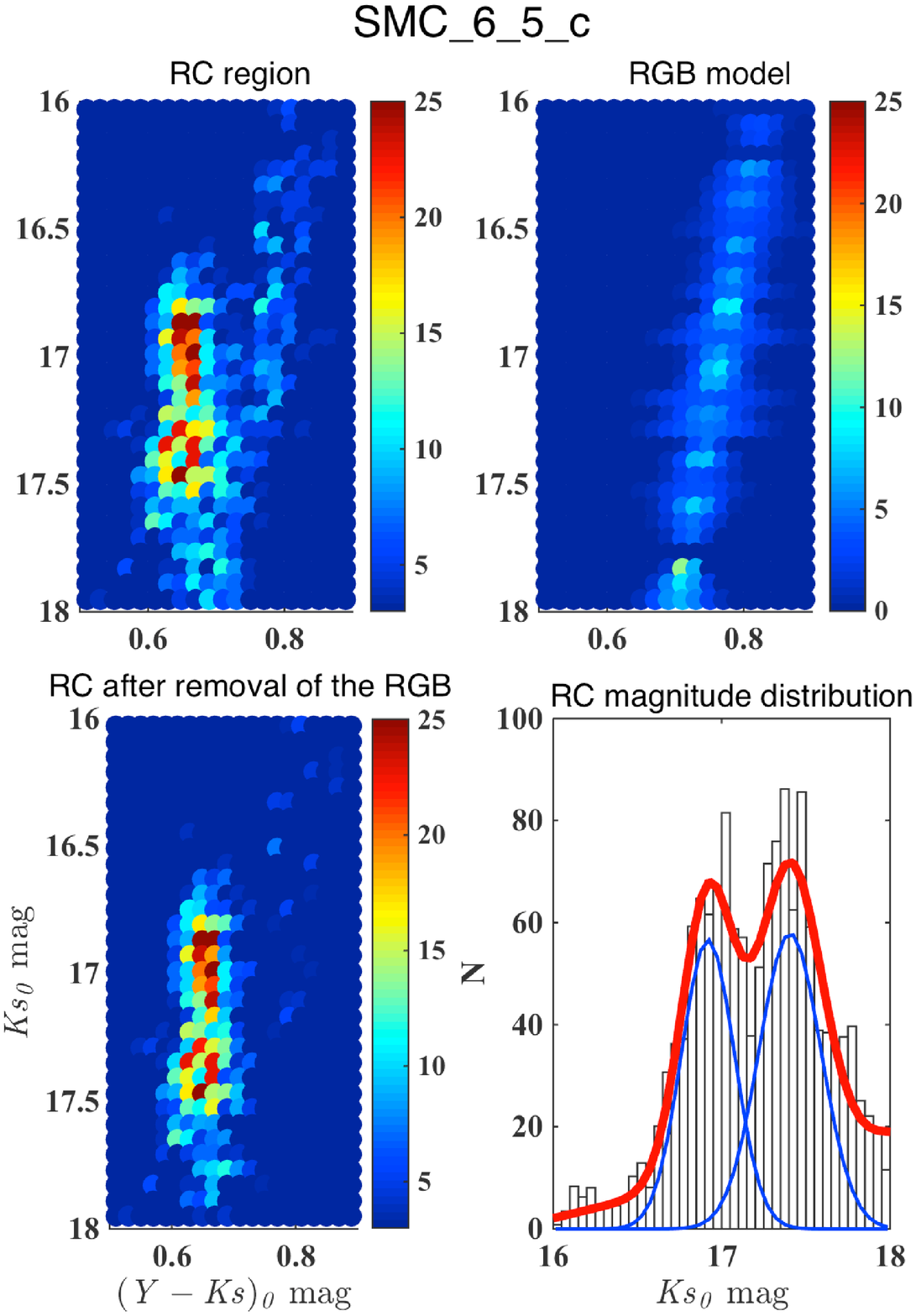} 
 &
 \hspace{-0.85cm}
\includegraphics[height=0.55\textwidth,width=0.52\textwidth]{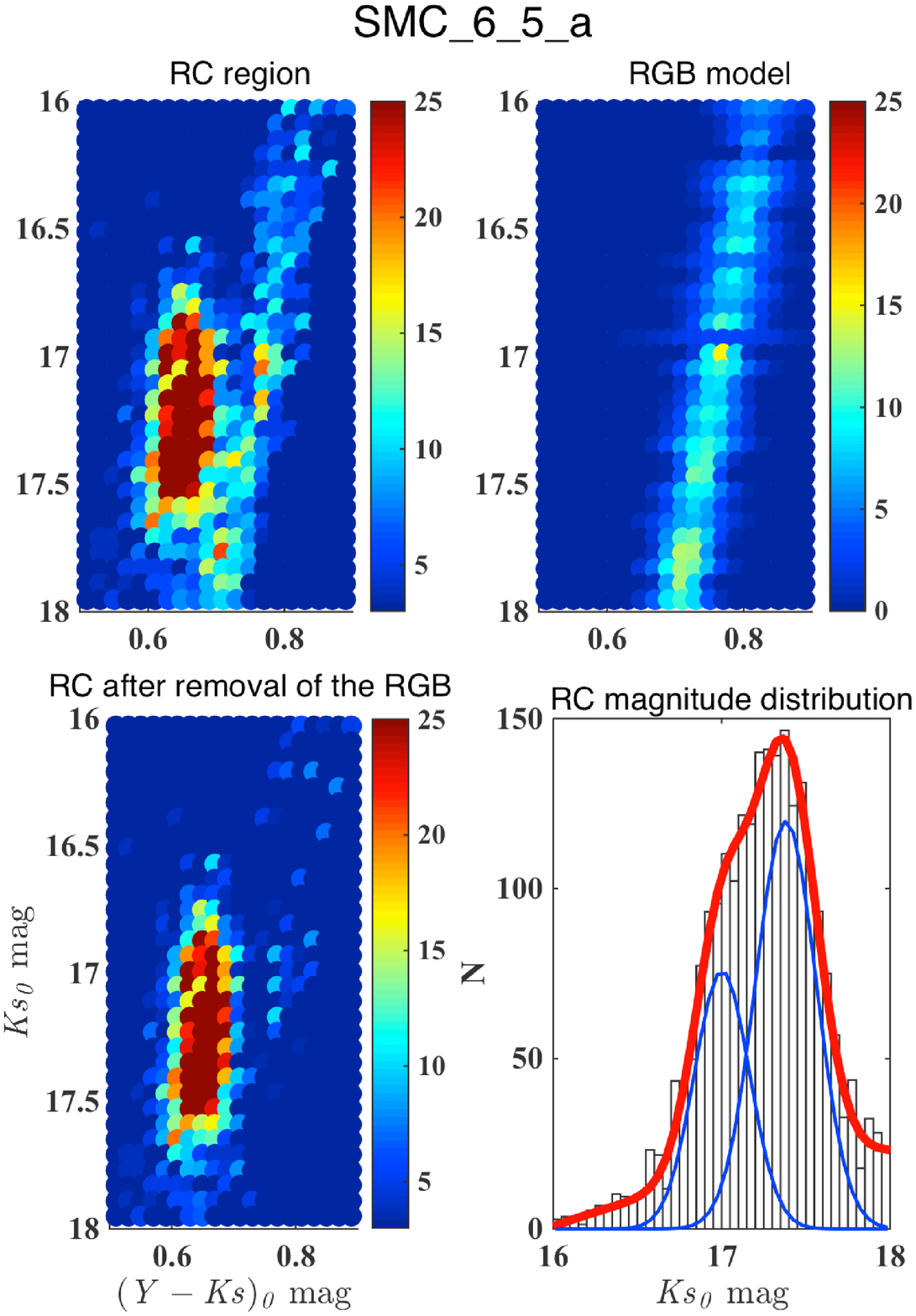} \\
\end{tabular}
\caption{Steps involved in deriving the luminosity function of the RC stars and the profile fitting to the distribution for the four sub-fields of tile SMC 6\_5. The top-left and top-right plots in each panel show the reddening-corrected and MW-foreground-subtracted RC region and the RGB model obtained for the sub-field respectively. The bottom-left and bottom-right plots in each panel show the RGB subtracted RC region and the luminosity function of the RC stars with the profile fits for the sub-field respectively. The total fit (thick-red line) to the distribution and the separate components (thin-blue line) are also shown in the bottom-right plot.}
\end{figure*}

\begin{table*}
 \caption{Gaussian parameters and the reduced $\chi^2$ values of the profile fits to the luminosity function of the RC stars in the sub-regions.}
 \centering
 \large
 \begin{tabular}{crccrccc}
\hline
Sub-region & $a_0$ & $a_1$ & $a_2$ & $b_0$ & $b_1$ & $b_2$&$\chi^2$\\
&Height&Centre&Width&Height&Centre&Width&\\
\hline
SMC\_6\_5\_a & 119.9$\pm$11.1& 17.38$\pm$0.04&0.18$\pm$0.02 &75.8$\pm$16.2 &17.00$\pm$0.05 &0.16$\pm$0.02 &1.22\\
SMC\_6\_5\_b &49.2$\pm$6.1 &17.51$\pm$0.03 & 0.16$\pm$ 0.02&51.9$\pm$4.4 &17.04$\pm$0.04 &0.21$\pm$0.03 &1.27\\
SMC\_6\_5\_c &57.9$\pm$4.1 &17.41$\pm$0.02 &0.18$\pm$0.02 &56.6$\pm$4.4 &16.92$\pm$0.02&0.16$\pm$0.02&1.38 \\
SMC\_6\_5\_d & 42.8$\pm$4.3&17.44$\pm$0.03 &0.15$\pm$0.02 &59.5$\pm$4.1 &16.97$\pm$0.02 &0.19$\pm$0.02&1.37\\
SMC\_6\_3\_a &188.3$\pm$18.4 &17.41$\pm$0.01 &0.12$\pm$0.01 &56.88$\pm$16.7 &17.41$\pm$ 0.05&0.31$\pm$0.06 &1.16\\
SMC\_6\_3\_b &78.8$\pm$14.6 &17.44$\pm$0.01 &0.12$\pm$0.02 &52.5$\pm$13.6 &17.43$\pm$0.04 &0.29$\pm$0.05 &1.02\\
SMC\_6\_3\_c &208.8$\pm$31.1 &17.38$\pm$0.01 &0.14$\pm$0.01 &65.1$\pm$28.6 &17.24$\pm$0.09 &0.26$\pm$0.04 &1.27\\
SMC\_6\_3\_d & 96.6$\pm$11.1&17.4$\pm$0.01 &0.10$\pm$0.01 &56.8$\pm$9.6 &17.36$\pm$0.03 &0.29$\pm$0.04&1.48\\
SMC\_5\_6\_a &46.3$\pm$4.7 &17.41$\pm$0.02 &0.13$\pm$0.02 &51.8$\pm$4.2 &16.95$\pm$0.02 &0.18$\pm$0.02&0.88\\
SMC\_5\_6\_b & 34.5$\pm$3.3&17.42$\pm$0.03 &0.18$\pm$0.03 &53.6$\pm$3.8 &16.92$\pm$0.02 &0.18$\pm$0.02&1.11\\
SMC\_5\_6\_c &12.3$\pm$2.7 &17.40$\pm$0.08 &0.17$\pm$0.07 &39.2$\pm$3.5 &16.93$\pm$0.03 &0.17$\pm$0.02&1.03\\
SMC\_5\_6\_d &16.6$\pm$3.7 &17.31$\pm$0.03 &0.10$\pm$0.03 &37.8$\pm$3.5 &16.90$\pm$0.02 &0.16$\pm$0.02&1.18\\
SMC\_5\_5\_a & 361.3$\pm$9.5& 17.29$\pm$0.01& 0.19$\pm$0.01&57.9$\pm$9.4 &16.87$\pm$0.02&0.09$\pm$0.02 &1.28 \\
SMC\_5\_5\_b &210.7$\pm$6.9 &17.28$\pm$0.01 &0.21$\pm$0.01 &42.8$\pm$10.0 &16.86$\pm$0.03 &0.09$\pm$0.02 & 1.18\\
SMC\_5\_5\_c &92.3$\pm$7.7 &17.33$\pm$0.06 & 0.29$\pm$0.06& 35.02$\pm$22.7&16.87$\pm$0.07 &0.17$\pm$0.05 &1.06\\
SMC\_5\_5\_d & 80.3$\pm$4.8& 17.38$\pm$0.03&0.19$\pm$0.02 &66.4$\pm$7.1 &16.95$\pm$0.03 &0.14$\pm$0.02 & 1.06\\
SMC\_5\_4\_a & 1347$\pm$68.5&17.29$\pm$0.01 &0.10$\pm$0.01 &1030.2$\pm$67.1 &17.23$\pm$0.01 &0.23$\pm$0.01 &1.58 \\
SMC\_5\_4\_b & 610.4$\pm$44.4&17.29$\pm$0.01 &0.10$\pm$0.01 &522.5$\pm$43.4 &17.22$\pm$0.01 &0.23$\pm$0.01 & 1.47\\
SMC\_5\_4\_c & 598.7$\pm$46.3& 17.30$\pm$0.01& 0.11$\pm$0.01& 438.9$\pm$45.4& 17.25$\pm$0.01&0.25$\pm$0.01 & 2.53\\
SMC\_5\_4\_d &313.8$\pm$38.4 &17.27$\pm$0.01 &0.12$\pm$0.01 &262.9$\pm$35.9 &17.22$\pm$0.01 &0.28$\pm$0.02 & 1.34\\
SMC\_5\_3\_a & 907.4$\pm$53.7& 17.38$\pm$0.01& 0.11$\pm$0.01&462.3$\pm$58.8 &17.31$\pm$0.01 &0.23$\pm$0.01 & 1.01\\
SMC\_5\_3\_b & 364.1$\pm$27.2& 17.43$\pm$0.01&0.10$\pm$0.01 &223.4$\pm$26.6 &17.36$\pm$0.01 &0.23$\pm$0.01 & 3.82\\
SMC\_5\_3\_c &1253.6$\pm$32.3 &17.31$\pm$0.01 &0.14$\pm$0.01 &278.4$\pm$30.6 &17.19$\pm$0.03 &0.34$\pm$0.02 & 1.49\\
SMC\_5\_3\_d &430.9$\pm$58.9 &17.36$\pm$0.01 &0.10$\pm$0.01 &424.4$\pm$60.8 &17.30$\pm$0.01 &0.21$\pm$0.01 & 1.10\\
SMC\_4\_5\_a & 271.7$\pm$11.1& 17.38$\pm$0.01&0.15$\pm$0.01 &74.7$\pm$7.9 &16.95$\pm$0.05 &0.19$\pm$0.03 &1.17\\
SMC\_4\_5\_b &328.9$\pm$42.9 & 17.33$\pm$0.03& 0.18$\pm$0.01& 69.2$\pm$36.2&16.96$\pm$0.17 &0.21$\pm$0.07 & 1.18\\
SMC\_4\_5\_c & 60.6$\pm$3.9& 17.38$\pm$0.02&0.23$\pm$0.03 &31.3$\pm$5.3 &16.87$\pm$0.03 &0.14$\pm$0.02 &1.13\\
SMC\_4\_5\_d & 74.9$\pm$6.2& 17.38$\pm$0.04&0.19$\pm$0.03 &55.3$\pm$8.9 &16.94$\pm$0.05 &0.18$\pm$0.03 &1.12\\
SMC\_4\_4\_a & 1491.6$\pm$68.8&17.34$\pm$0.01 &0.09$\pm$0.01 &1471.3$\pm$0.01 &17.27$\pm$0.01 &0.22$\pm$0.01 & 1.32\\
SMC\_4\_4\_b & 2288.2$\pm$91.3& 17.30$\pm$0.01& 0.12$\pm$0.01& 1055.1$\pm$80.8&17.27$\pm$0.01 &0.29$\pm$0.02 & 1.61\\
SMC\_4\_4\_c & 602.4$\pm$38.4&17.34$\pm$0.01 &0.11$\pm$0.01 &275.4$\pm$35.4 &17.28$\pm$0.02 &0.27$\pm$0.02 & 1.77\\
SMC\_4\_4\_d &812.7$\pm$66.5 &17.28$\pm$0.01 &0.22$\pm$0.01 &847.9$\pm$67.9 &17.31$\pm$0.01 &0.10$\pm$0.01 &1.15\\
SMC\_4\_3\_a & 1441.2$\pm$105.5& 17.33$\pm$0.01& 0.11$\pm$0.01& 941.5$\pm$107.9& 17.29$\pm$0.01& 0.22$\pm$0.01&1.55 \\
SMC\_4\_3\_b & 1080.8$\pm$85.7& 17.33$\pm$0.01&0.11$\pm$0.01 &848.4$\pm$85.1 &17.29$\pm$0.01&0.24$\pm$0.01&1.38 \\
SMC\_4\_3\_c &1769.1$\pm$47.9 &17.30$\pm$0.01 &0.13$\pm$0.01 &584.9$\pm$44.6&17.20$\pm$0.01 & 0.34$\pm$0.02&1.47\\
SMC\_4\_3\_d & 1615.3$\pm$41.1& 17.29$\pm$0.01& 0.13$\pm$0.01& 797.2$\pm$47.6&17.24$\pm$0.02 &0.42$\pm$0.03 &1.24\\
SMC\_4\_2\_a & 306.1$\pm$8.7& 17.41$\pm$0.01& 0.16$\pm$0.01& -& -& -&1.54\\
SMC\_4\_2\_b & 279.9$\pm$8.5&17.43$\pm$0.01 &0.15$\pm$0.01 &- &- &-&1.72\\
SMC\_4\_2\_c &611.4$\pm$53.5 &17.39$\pm$0.01 &0.13$\pm$0.01 &245.3$\pm$55.1 &17.33$\pm$0.02 &0.26$\pm$0.02&1.71\\
SMC\_4\_2\_d &486.9$\pm$54.3 &17.42$\pm$0.01 &0.13$\pm$0.01 &202.3$\pm$54.9&17.32$\pm$0.03 &0.22$\pm$0.01 &1.75\\
SMC\_3\_5\_a &117.9$\pm$6.3 &17.39$\pm$0.02 &0.18$\pm$0.02 & 63.9$\pm$5.5& 16.90$\pm$0.03&0.19$\pm$0.03&0.97\\
SMC\_3\_5\_b &153.5$\pm$6.1 &17.35$\pm$0.01 &0.20$\pm$0.01 & 50.1$\pm$6.4& 16.85$\pm$0.03&0.16$\pm$0.02&1.30\\
SMC\_3\_5\_c &48.6$\pm$4.2 &17.4$\pm$0.02 &0.13$\pm$0.01 &25.4$\pm$2.7 &16.89$\pm$0.03 &0.21$\pm$0.04&1.15\\
SMC\_3\_5\_d & 59.9$\pm$4.9& 17.37$\pm$0.01&0.15$\pm$0.01 &31.7$\pm$3.9 &16.89$\pm$0.02 &0.14$\pm$0.02&0.99\\
SMC\_3\_3\_a & 696.8$\pm$86.1& 17.40$\pm$0.01& 0.14$\pm$0.01& 105.7$\pm$56.0&17.18$\pm$0.17 &0.24$\pm$0.06 &2.02 \\
SMC\_3\_3\_b &1214.0$\pm$100.6 &17.37$\pm$0.01 &0.11$\pm$0.01 & 712.6$\pm$104.6&17.30$\pm$0.01 &0.21$\pm$0.01 & 1.18\\
SMC\_3\_3\_c &763.4$\pm$90.6 &17.37$\pm$0.01 &0.15$\pm$0.01 &130.4$\pm$49.1 &17.08$\pm$0.16 &0.26$\pm$0.07&2.68 \\
SMC\_3\_3\_d & 1385.2$\pm$57.8&17.35$\pm$0.01 &0.10$\pm$0.01 &1031.0$\pm$55.4 &17.29$\pm$0.01 &0.25$\pm$0.01 & 1.30\\
SMC\_3\_1\_a+c & 69.7$\pm$4.0& 17.38$\pm$0.01& 0.18$\pm$0.01& -& -&-&1.08\\
SMC\_3\_1\_b+d &113.3$\pm$5.7 &17.36$\pm$0.01 &0.14$\pm$0.01 &- &- &-&1.33\\

\hline
\end{tabular}
 
 \end{table*}

\begin{landscape}
\begin{figure}
\centering
\setlength{\tabcolsep}{1pt}
\renewcommand{\arraystretch}{0.00001}
\begin{tabular}{ccccccc}
&
\includegraphics[height=0.25\textwidth,width=0.22\textwidth]{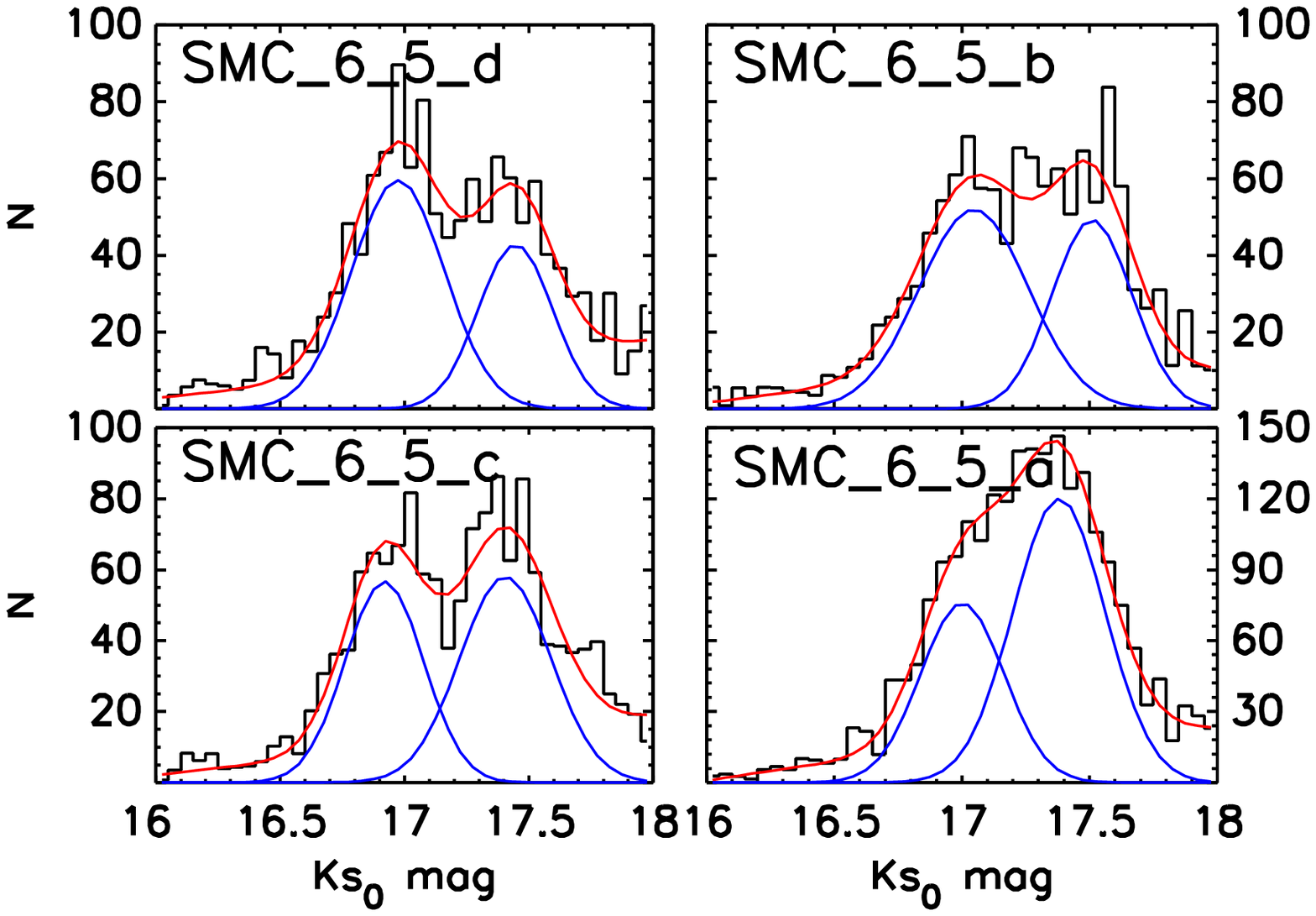} &  
 &
\includegraphics[height=0.25\textwidth,width=0.22\textwidth]{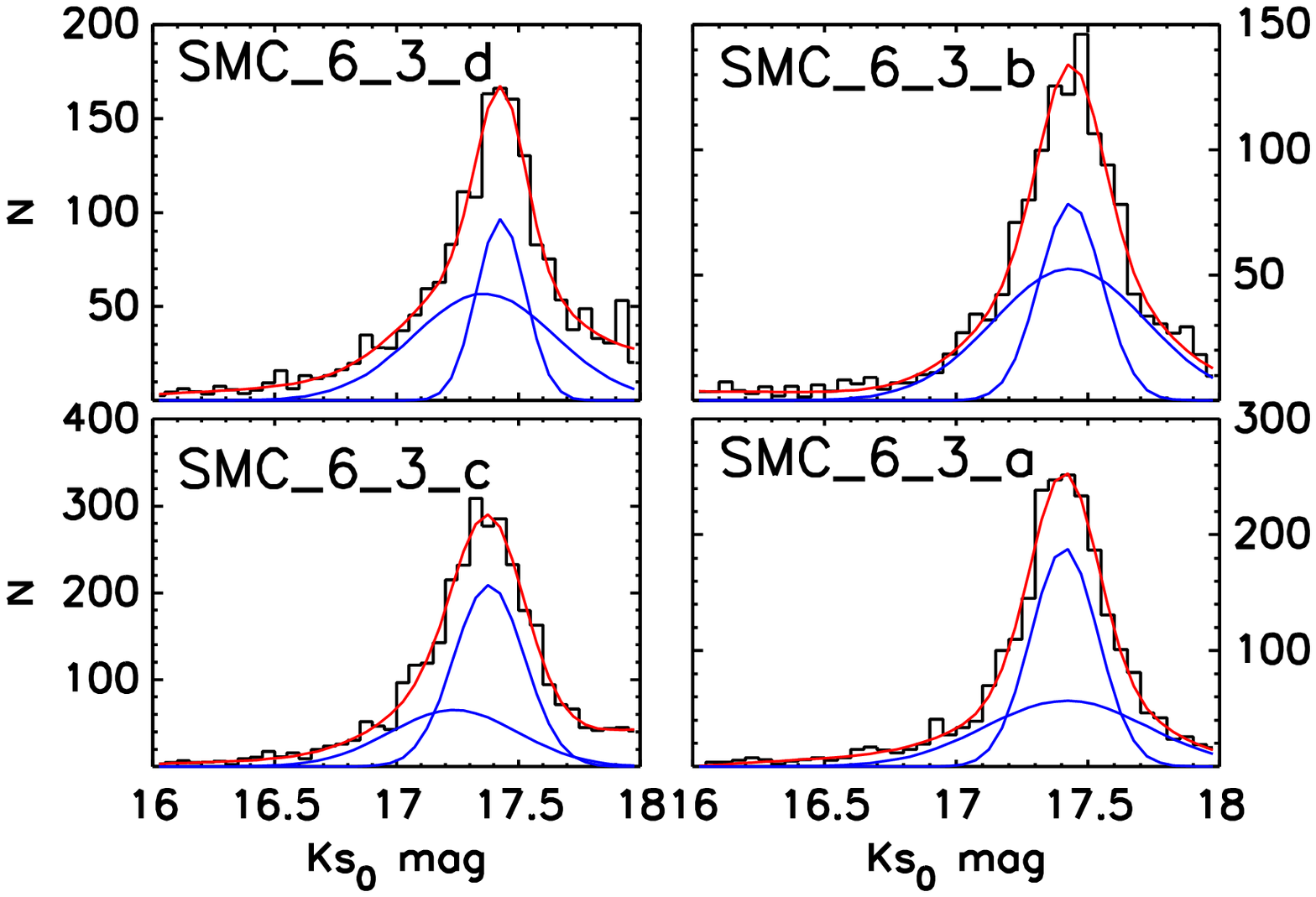}&
&
&
\\
\includegraphics[height=0.25\textwidth,width=0.22\textwidth]{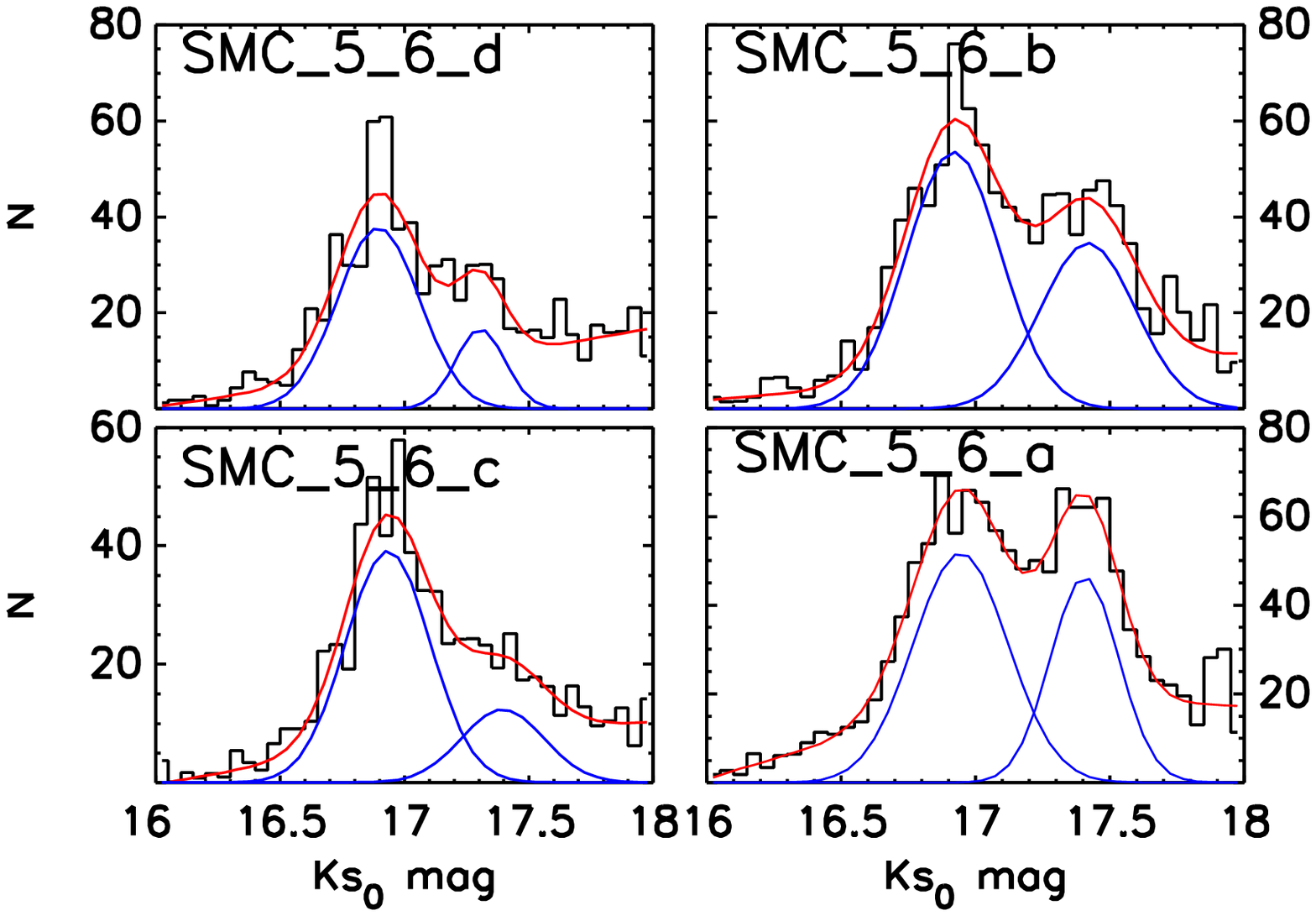}&
\includegraphics[height=0.25\textwidth,width=0.22\textwidth]{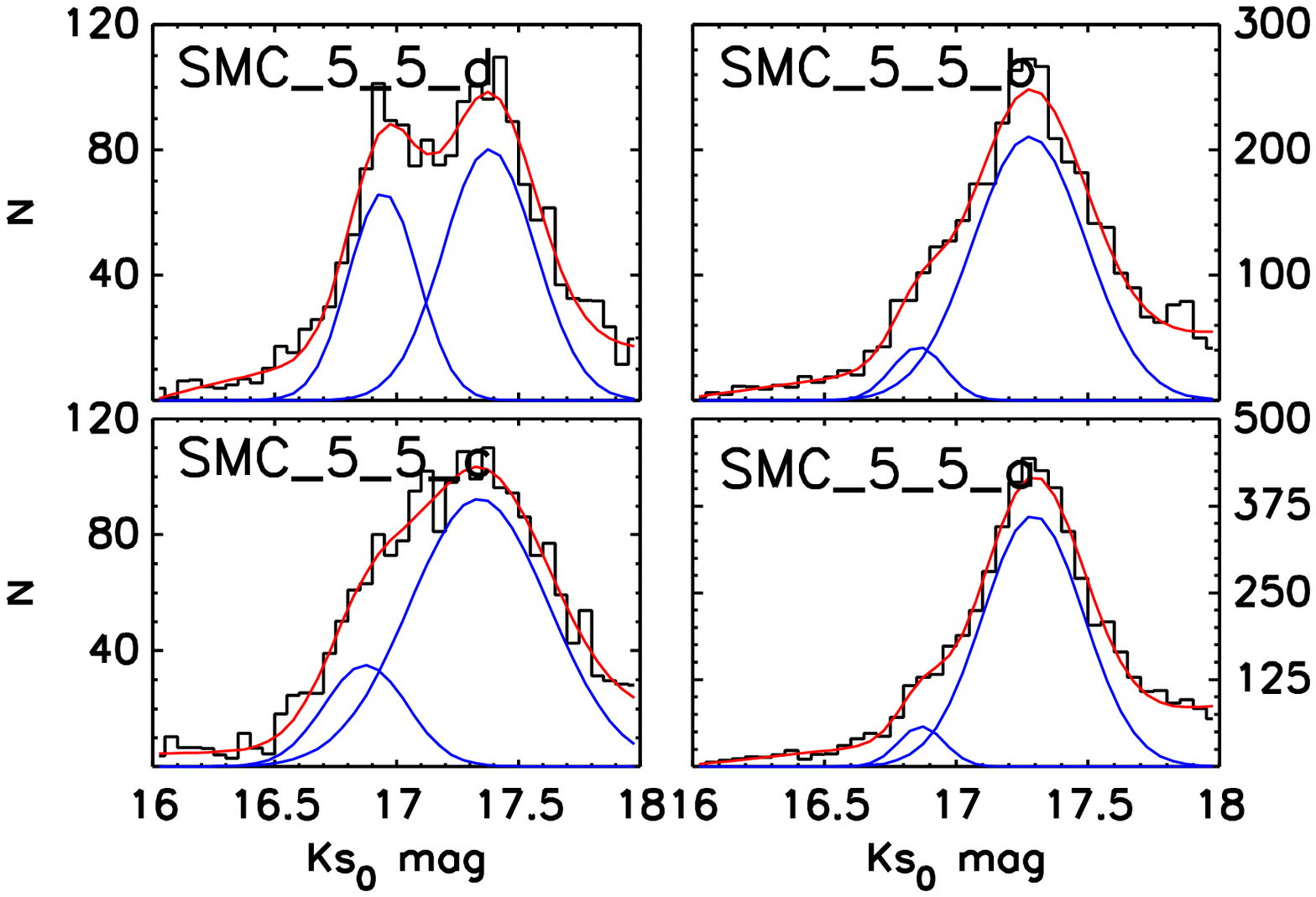}&
\includegraphics[height=0.25\textwidth,width=0.22\textwidth]{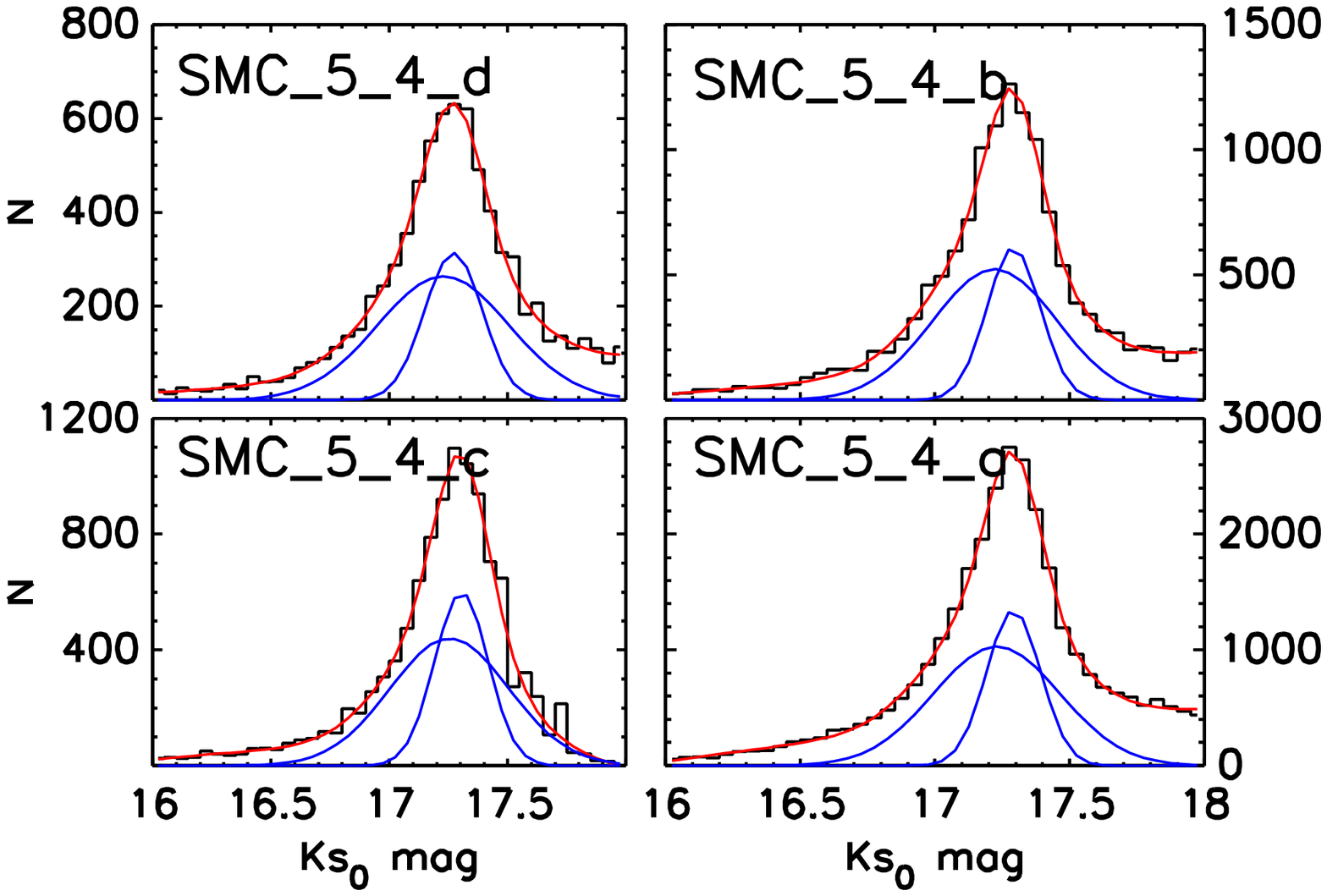}&
\includegraphics[height=0.25\textwidth,width=0.22\textwidth]{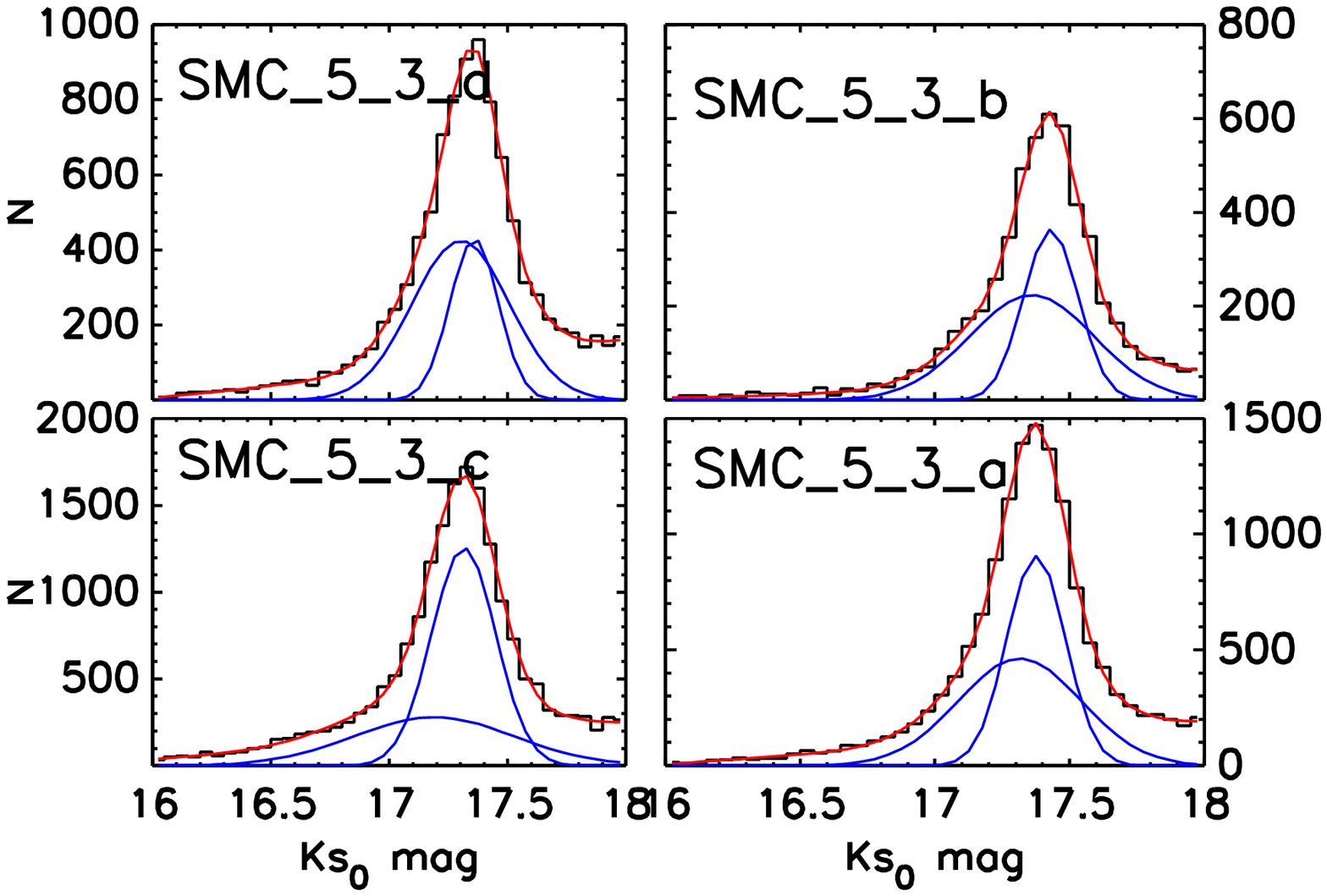}&
&
&\\

&
\includegraphics[height=0.25\textwidth,width=0.22\textwidth]{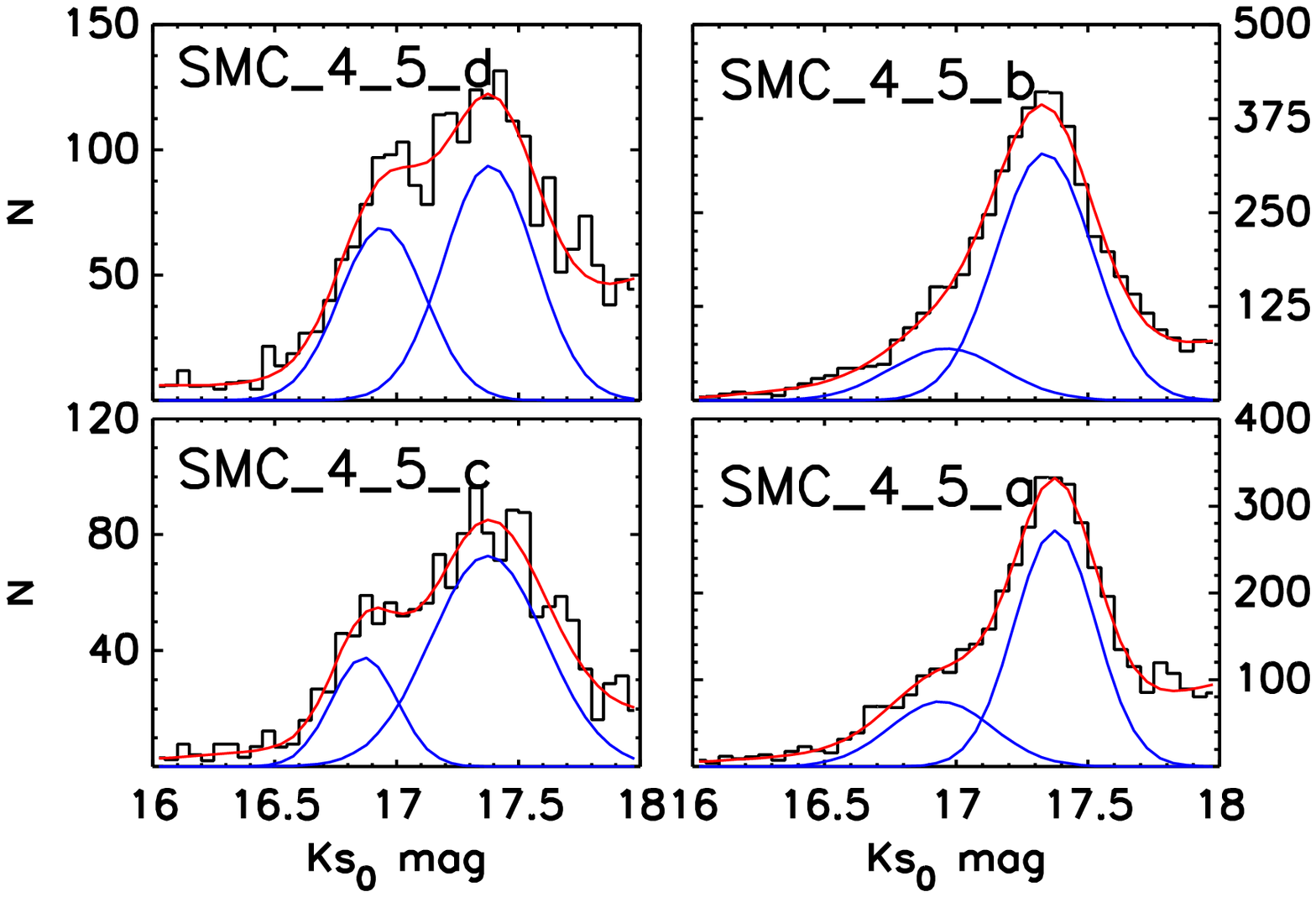}&
\includegraphics[height=0.25\textwidth,width=0.22\textwidth]{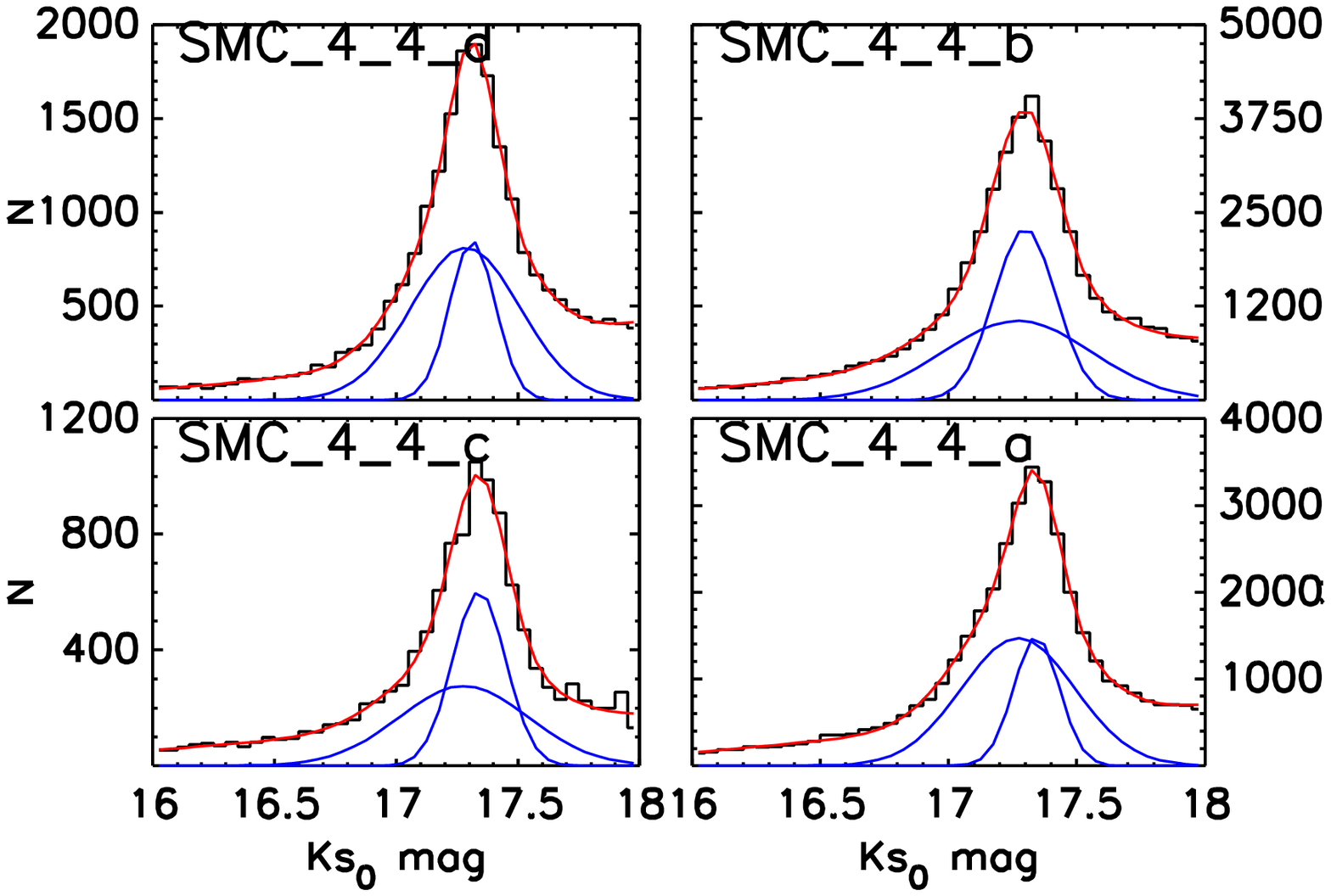}&
\includegraphics[height=0.25\textwidth,width=0.22\textwidth]{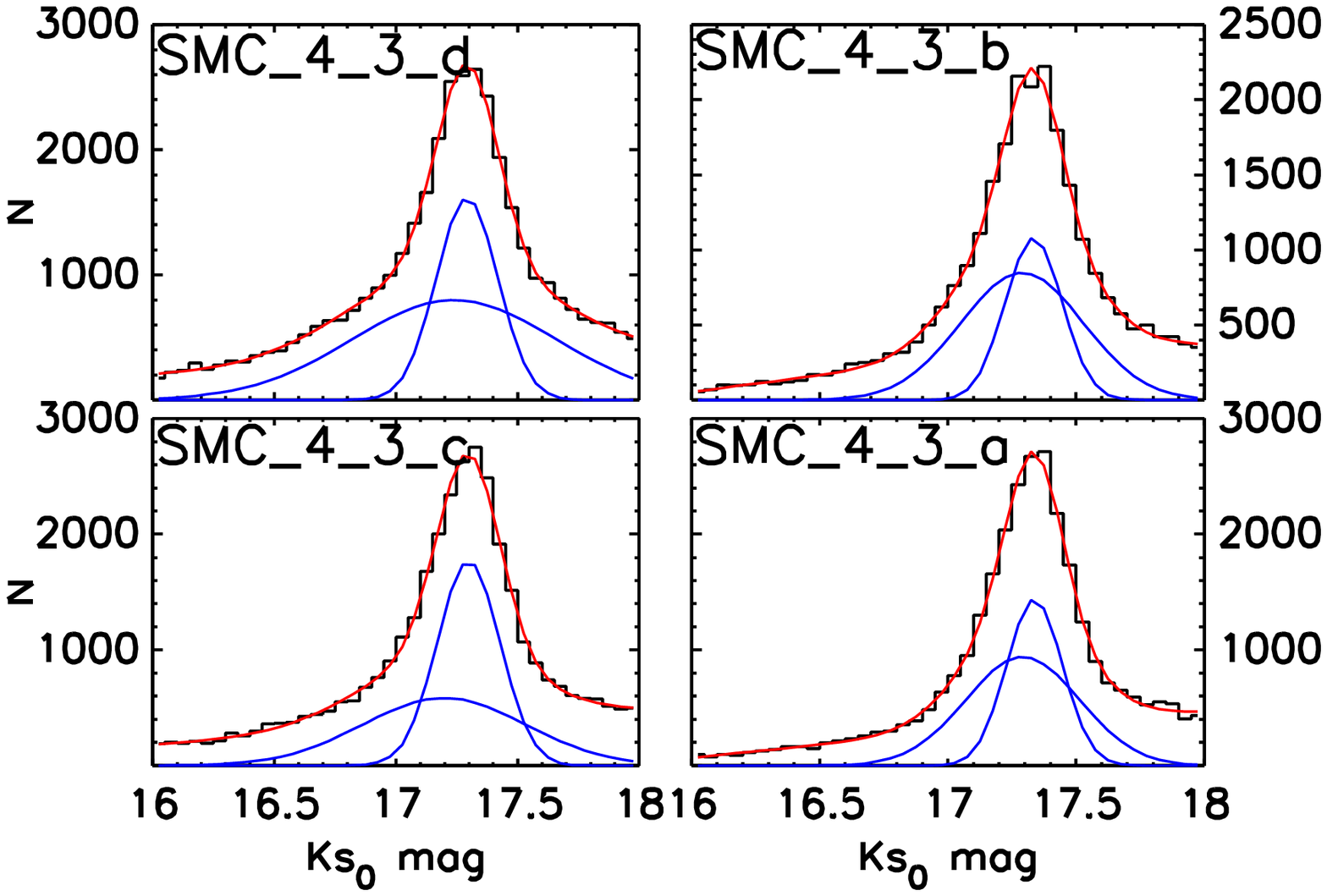}&
\includegraphics[height=0.25\textwidth,width=0.22\textwidth]{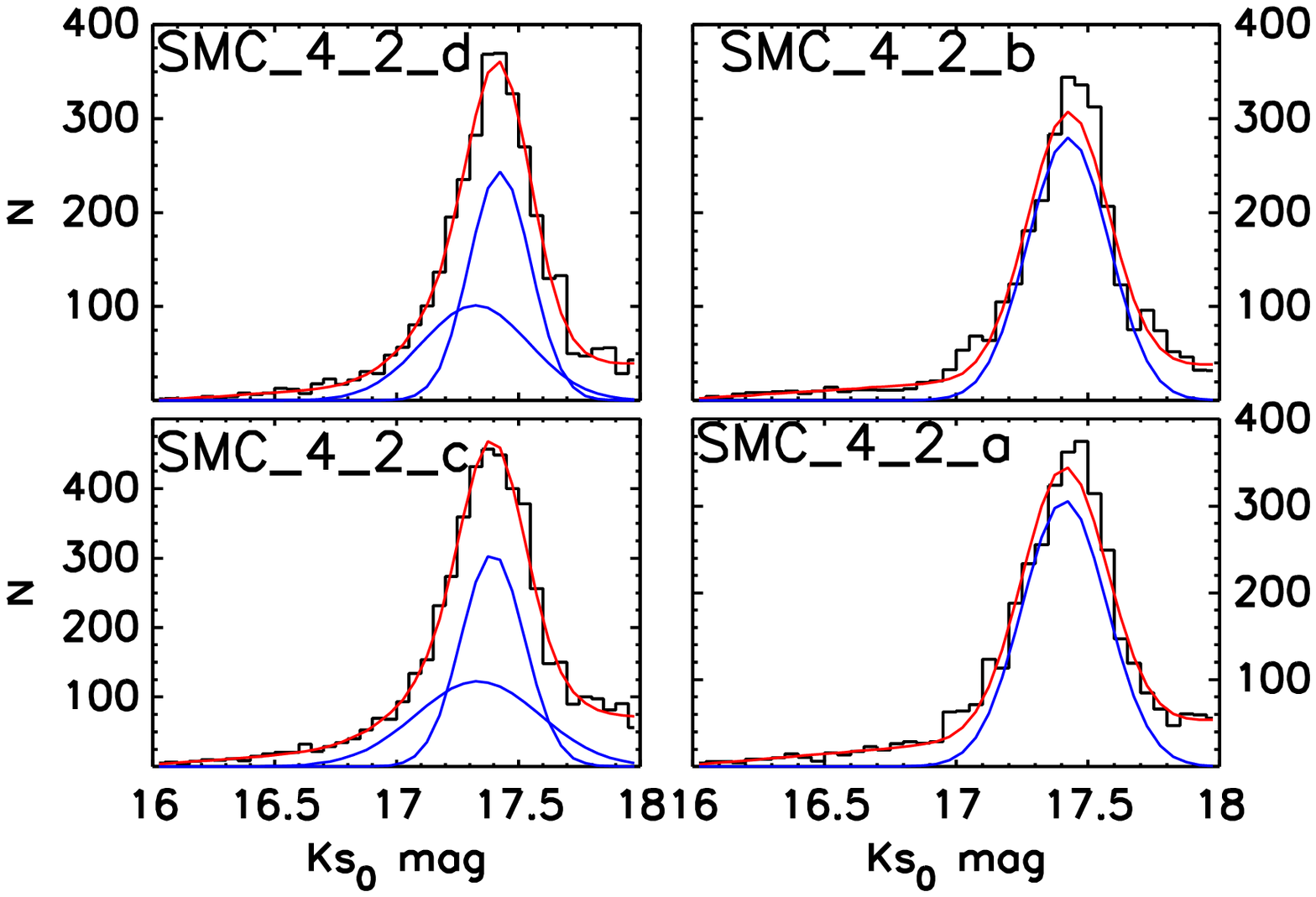} &
&\\

&
\includegraphics[height=0.25\textwidth,width=0.22\textwidth]{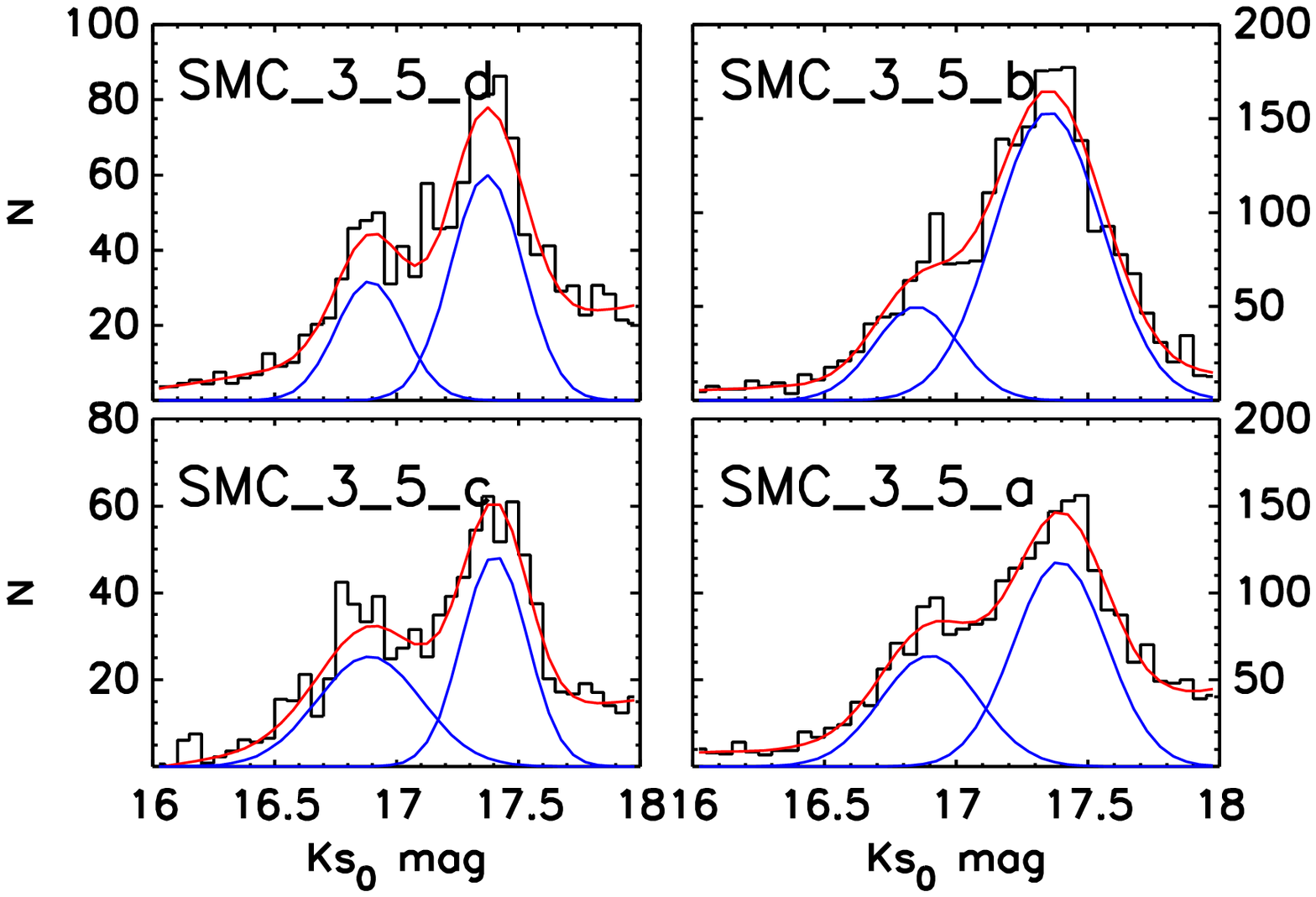}&
&
\includegraphics[height=0.25\textwidth,width=0.22\textwidth]{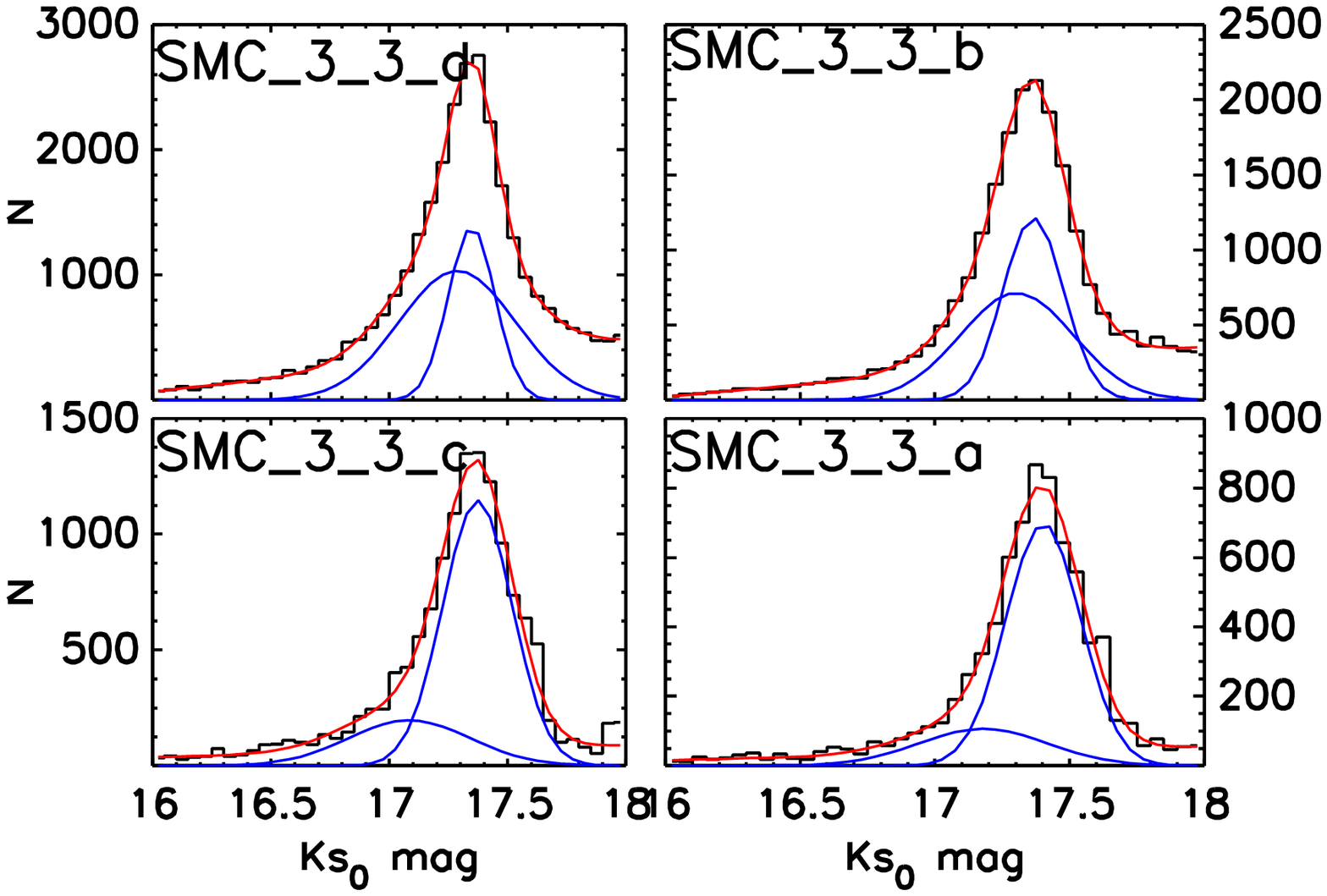} &
&
\includegraphics[height=0.25\textwidth,width=0.22\textwidth]{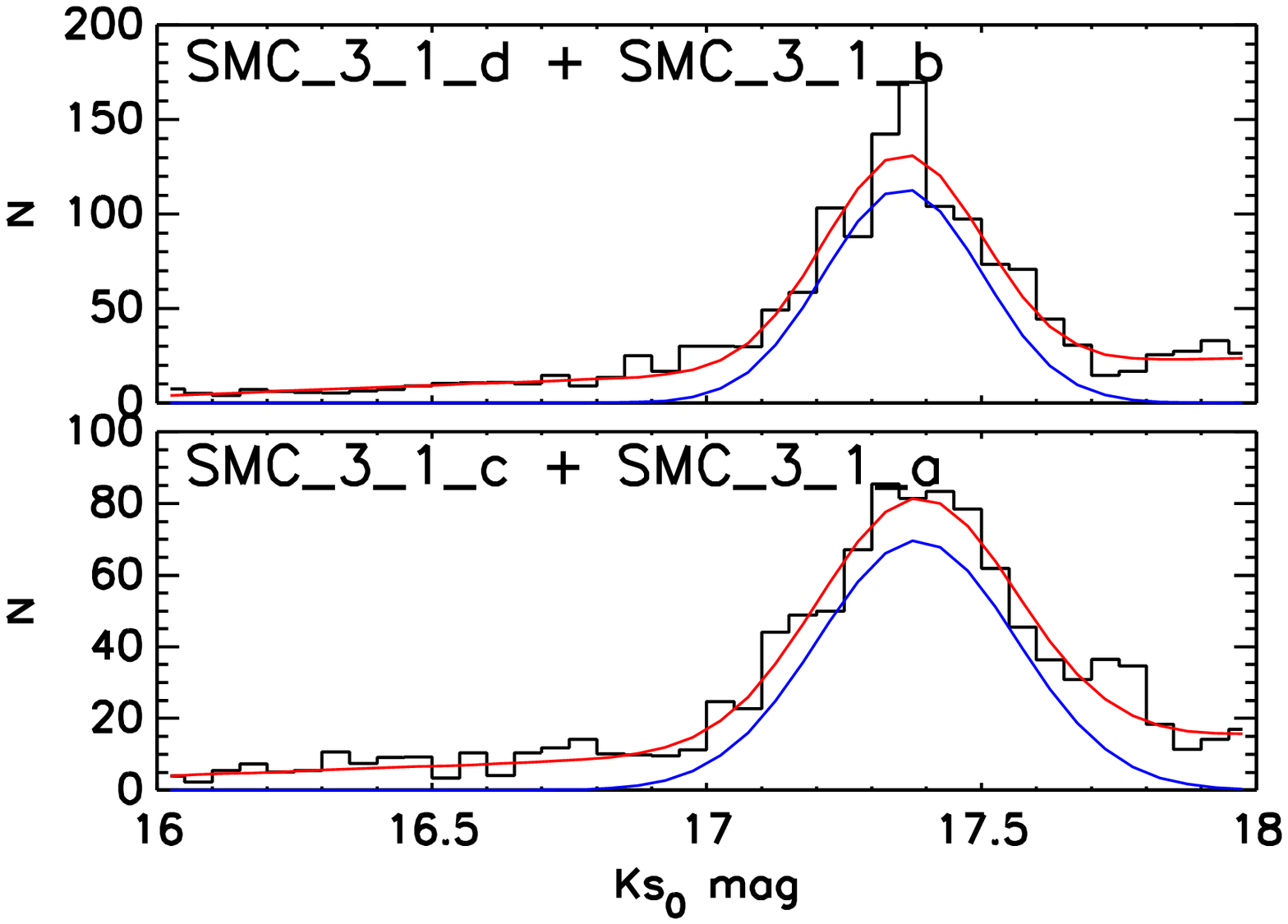}\\

\end{tabular}
\caption{Luminosity function of the RC stars in different sub-regions. Each panel shows the luminosity function of the sub-fields in the SMC tiles. The black histograms show the observed luminosity functions and the red lines show the total fits to the distributions. The blue lines represent the separate components of the fits.}
\end{figure}
\end{landscape}

\section{Cause of bimodality: Distance effect} 

Since RC stars are standard candles, the natural explanation for the observed  bimodality in their  luminosity function is a distance effect.  In all five eastern tiles from $\sim$ 2$^\circ$.5 $\le$ $\it{r}$ $\le$ 4$^\circ$.0, there is one peak at  $K_\textnormal{s,0}$ $\sim$ 17.28 -- 17.51 mag (average = 17.38 $\pm$ 0.05 mag) and a brighter peak at $K_\textnormal{s,0}$ $\sim$ 16.87 -- 17.04 mag (average = 16.92 $\pm$ 0.05 mag). A difference of 0.46 mag in the average magnitudes of the brighter and fainter clumps corresponds to a distance variation of $\sim$ 11.8 $\pm$ 2.0 kpc, if we assume the faint clump is at the distance of the main body of the SMC. The central tiles show a broad component (width of $\sim$ 0.25 -- 0.4 mag) with a slightly brighter peak, 0.05 -- 0.15 mag brighter relative to the peak of the narrow component, at $K_\textnormal{s,0}$ $\sim$ 17.36 $\pm$ 0.05 mag. 

\subsection{Arguments against a single distance}

An extinction of 0.46 mag in the $\it{Ks}$ band should produce a colour difference of 1.1 mag in $(Y-K_\textnormal{s})$ colour. As can be appreciated from Fig. 4, the colour difference between the brighter and fainter clumps is minimal and, hence, the effect of extinction in the observed bimodality in the eastern tiles is very small. There could be an effect of differential extinction within a tile and this could contribute to the width of the RC luminosity function, especially in the central tiles. But the effect of extinction in NIR bands is minimal and  \cite{Rubele2015} suggest a differential extinction of $A_{Ks}$ $\sim$ 0.01 mag. In order to check this further, the colour distributions of the RC (after subtracting the RGB) in the sub-regions are analysed. The RC colour distribution of four sub-regions (SMC 4\_3\_d, SMC 6\_3\_d, SMC 5\_6\_d and SMC 3\_5\_d), which are at different locations in the SMC, are shown in Figure 7. We see that they can be approximated by a Gaussian distribution in the $(Y-K_\textnormal{s})_0$ colour range 0.55 -- 0.85 mag (the stars in this range are used for the construction of RC luminosity function).  Gaussian fits to the distributions provide dispersions in the RC colour distributions, which are measures of the differential reddening.  The dispersion values obtained for the sub-regions are in the range, 0.03 -- 0.07 mag, with the highest in the central regions. All the sub-regions except those in the central tiles (SMC 4\_3, SMC 4\_4 and SMC 5\_3), have dispersions on the order of the photometric error in the $(Y-K_\textnormal{s})_0$ colour (in the magnitude range of the RC, the photometric error in $(Y-K_\textnormal{s})_0$ is $\sim$ 0.03 mag). After subtracting the contribution from photometric errors, the central tiles could have a contribution from differential reddening amounting to $E(Y-Ks)$ $\sim$ 0.05 -- 0.06 mag. This corresponds to a differential extinction of $A_{Ks}$ $\sim$ 0.03 mag. This extinction could not explain the observed width of the broad components in the central tiles nor the bimodality in the eastern tiles. However, the dispersion in RC colour could also have some contribution from the intrinsic spread of RC stars due to population effects. Thus the differential extinction of $A_{Ks}$ $\sim$ 0.03 mag is an upper limit.

\begin{figure}
\centering
 \includegraphics[width=0.5\textwidth]{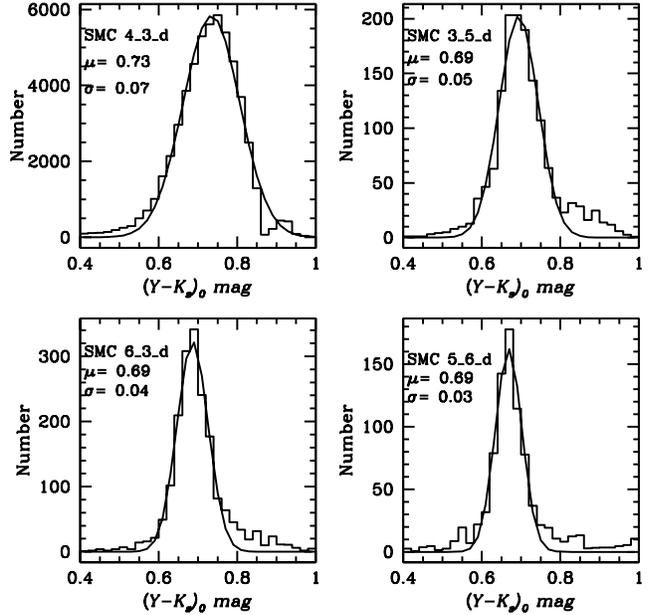} 
 \caption{Colour distributions of RC stars in four sub-regions. The profile fit and the Gaussian parameters are also shown.}
\end{figure}

Another important point to consider is that in the SMC areas where these RC structures appear as two clearly distinct clumps, the fainter clump nearly coincides in brightness with the narrow component of the RC observed in the SMC centre. This indicates that the fainter clump corresponds to the classical intermediate-age RC (age of $\sim$ 2 -- 9 Gyr) observed at the same $\sim$ 60 kpc distance as the SMC's main body. In galaxy regions with ~1-Gyr old populations of intermediate to high metallicities, population effects can easily cause the appearance of an extension of this classical RC towards fainter magnitudes, which is the so-called secondary red clump \citep{Girardi1999}, but this is not what is observed here. To explain the presence of a RC brighter than the intermediate-age RC, one has to resort to populations of even younger ages (of, say, ~0.5~Gyr; \citealt{Ripepi2014}), but then the difficulties are of another kind: As illustrated in \cite{Girardi2009} and in fig. 5 of \cite{Girardi2016}, the RC of younger populations are much more elongated in brightness than the intermediate-age RC (with an r.m.s. dispersion of 0.25 mag compared with 0.1 mag for the faint clump). Even if a short-lived recent burst of star formation could cause a bright RC appearing, on average, 0.46 mag brighter than the intermediate-age RC, it would be much broader ($\sim$ 2--3 times) than the faint clump. As can be appreciated from Table 1, the width of the bright and faint clumps in the eastern tiles are similar. To produce two RCs of similar spread in brightness, as observed in the eastern tiles of the SMC, one has to resort to more dramatic effects than a spread in age, including for instance the presence of a second population with a very different chemical composition (e.g., \citealt{Massari2014}; \citealt{Lee2015}), or a change in distance, which we explore in the next subsection. Recently, \cite{Lee2015} showed that a double RC feature (with a magnitude difference of 0.5 mag) in the Galactic bulge can be explained by multiple populations, where the brighter RC is formed from a helium enhanced second generation stars. This scenario is only valid in the metal rich ([Fe/H] $>$ $-0.1$, refer fig. 1 of \citealt{Lee2015}) regime and hence it is very unlikely to explain the double RC feature observed in the low metallicity (Fe/H $\sim$ $-$0.99 $\pm$ 0.01 dex, \citealt{Dobbie2014}) environment of the SMC.
 
Similarly, the population effects could contribute to the dispersion of the RC luminosity function, especially in the central tiles which show a broad component. Based on theoretical modelling, the 1$\sigma$ spread in the absolute K$_\textnormal{s}$ band luminosity function of the classical intermediate-age RC stars (ages of 2 -- 9 Gyr) is $\sim$ 0.05 -- 0.1 mag \citep{Girardi2016}. This spread along with differential extinction (0.03 mag) and photometric errors (0.03 mag) could only account for a dispersion of $\sim$ 0.07 -- 0.11 mag in the observed RC luminosity function. The observed width of the broad component in the central tiles is higher (0.2 -- 0.4 mag, with an average of $\sim$ 0.25 mag) than this value. The line-of-sight depth value corresponding to the average 1$\sigma$ width of the broad components in the central tiles, after correcting for the intrinsic spread of classical intermediate-age RC stars, differential extinction and photometric errors, is $\sim$ 6 kpc. This value is comparable with the line-of-sight depth ($\sim$ 5 kpc) estimates from other tracers, like the RR Lyrae stars (\citealt{KH2012}; \citealt{SS2009, SS2012}; \citealt{Haschke2012}) and Cepheids (\citealt{Haschke2012}; \citealt{SS2015}). 

However, the central tiles could include the presence of afore mentioned fainter secondary RC ($\sim$1-Gyr and $\sim$0.4 mag fainter) and the brighter blue loop stars ($\sim$0.5-Gyr and $\sim$0.4 mag brighter with a broad distribution). The presence of these populations (with a range in magnitude, $\sim$1 mag) along with a compact classical RC, could partially contribute to the width of the broad component in these central tiles. Their contribution to the width depends on the fraction of these young stars and their chemical properties.  The sub-regions (e.g. SMC 4\_3\_c, SMC 4\_3\_d and SMC 5\_3\_c) which show a relatively greater broad component width than the average, could have a significant fraction of these younger ($<$ 1 Gyr) RC stars. 

The best way to account for the population effects of the RC is to model the observed RC luminosity function using stellar population models, including the local star formation rate and age -- metallicity relation obtained from a high resolution star formation history map of the SMC.  But all the studies that recover the detailed star formation history, including the recent one by \cite{Rubele2015}, using the VMC data, assume zero-depth to the SMC. Harris \& Zaritsky (2004) simulated stellar populations with a distance spread of $\pm$ 0.2 mag (corresponding to a 1$\sigma$ depth of 6 kpc) and recovered the star formation history with a zero-depth model, using the same techniques as in \cite{Rubele2015}. This indicates that the current techniques to recover the star formation history of the SMC based on the single distance assumptions cannot clearly disentangle the effects of line-of-sight depth and population effects. Therefore, the modelling of the observed RC luminosity function using the currently available star formation history results would not enable us to quantify the population effects of the RC in the SMC.

However, \cite{Rubele2015} suggest a large-line-of-sight of depth in the south-eastern tile, SMC 3\_5, based on the total distance intervals corresponding to the 68 \% confidence level of the best-fit distances (fig. 7 of \citealt{Rubele2015}). Similar variations in the distances were also observed in some of the sub-regions of tiles SMC 6\_5 and SMC 4\_5. These authors also observed that the tile, SMC 5\_6 is at a closer distance ($\sim$ 54 kpc) to us and the residual of the fit to this tile revealed a less populated RC at fainter magnitudes. For central tiles, fig. 7 of \cite{Rubele2015} shows a width in the range, 1 -- 5 kpc. A follow-up study of the star formation history including a distance distribution will provide a tool to disentangle the effects of line-of-sight depth and population effects, especially in the central regions of the SMC.

This suggests that the observed bimodality of the RC luminosity function in the eastern tiles are most likely owing to the presence of stellar populations at two distances (separated by $\sim$ 10 -- 12 kpc). The broad component in the central tiles could have a contribution from both line-of-sight and population effects. The average line-of-sight depth estimates (1$\sigma$ depth of $\sim$ 6 kpc) in the central regions, which are comparable to that obtained from other tracers, indicates the presence of some fraction of a foreground RC population in these regions as well, but closer to the main body. We mainly concentrate on the foreground population in the eastern tiles in the following sections.

\subsection{Single-Distance Model v.s. Double-Distance Model}

We attempt to model the observed ($K_\textnormal{s}$, $Y-K_\textnormal{s}$) CMD of tile SMC~5\_6 using single and double-distance models. We consider the observed CMD as linear combinations of `stellar partial models' (SPMs), which are simulated simple stellar populations covering small intervals of age and metallicity. Our modelling is similar to the method adopted by \citet{Kerber2009} and \citet{Rubele2012, Rubele2015} and is briefly summarised here.

\begin{figure*}
\centering
\includegraphics[scale=1.00,angle=0]{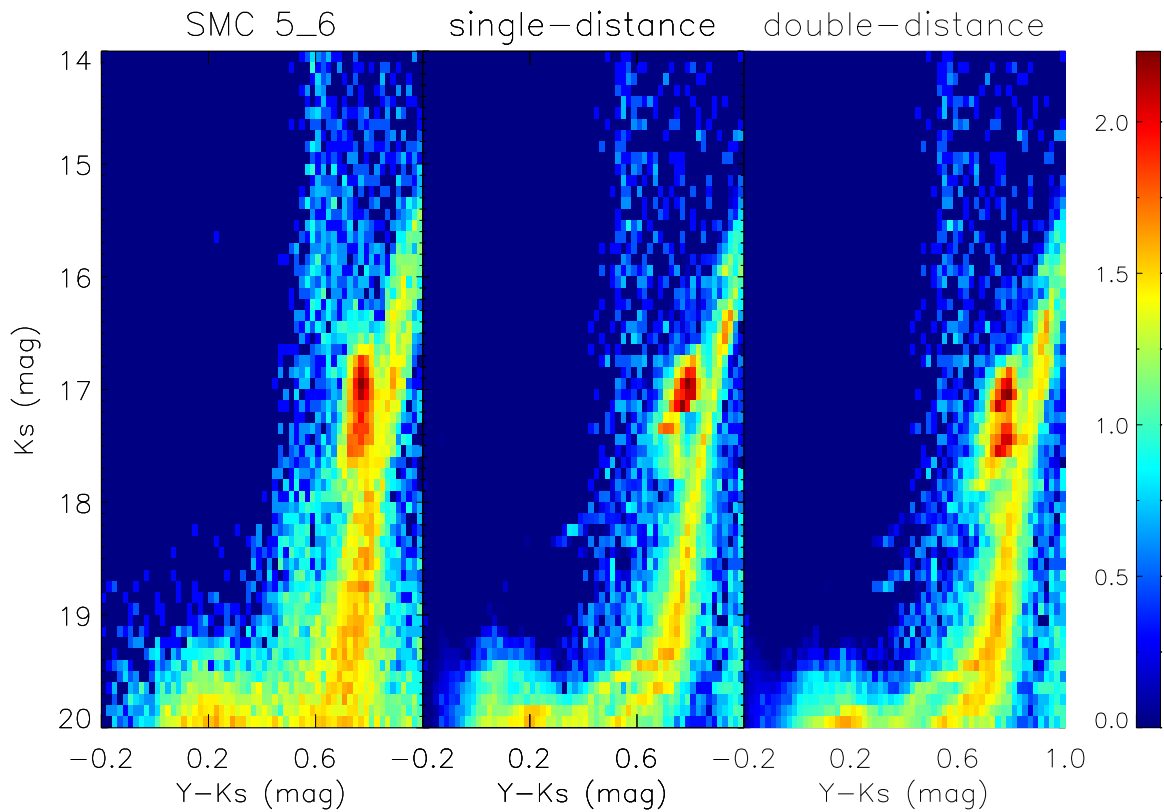}
\caption{Observed CMD for tile SMC~5\_6 (left-hand panel) and its best-fitting single-distance (middle panel) and double-distance (right-hand panel) models. The colour scale shows the logarithmic number of stars in each colour-magnitude bin.}
\label{d_fn.fig}
\end{figure*}

In the single-distance model, we use 14 SMC SPMs with ages from log($\tau$yr$^{-1}$)~=~6.9 to 10.075 and metallicities assigned according to the age -- metallicity relation of \citet{Piatti2011}. Unlike \citet{Kerber2009} and \citet{Rubele2012, Rubele2015}, who used 5 different SPMs for every age bin covering a significant range in metallicity, we simplify the problem by naively assuming that the stars strictly follow the age -- metallicity relation. The ages and metallicities of the SPMs are thus the first and fourth columns of Table~2 of \citet{Rubele2015}. The SPMs are simulated with the PARSEC~v1.2S stellar evolutionary tracks \citep{Bressan2012} while adopting a \citet{Chabrier2001} log-normal initial mass function and a 30\% binary fraction. The simulated binaries are non-interacting systems and have primary/secondary mass ratios evenly distributed from 0.7 to 1.0. Galactic foreground stars are simulated with TRILEGAL \citep{Leo2005} and are added as an additional SPM.

The SMC SPMs are displaced using a distance modulus $(m-M)_0$ and an extinction $A_V$.  We vary $(m-M)_0$ from 18.3 to 18.9 mag in steps of 0.05 mag and $A_V$ from 0.1 to 0.9 mag in steps of 0.05 mag. For each pair of $(m-M)_0$ and $A_V$, we convolve the SPMs with photometric errors and completeness. The CMD is divided into bins of $\Delta K_\textnormal{s}$~=~0.1~mag and $\Delta (Y - K_\textnormal{s})$~=~0.02~mag, with limits of 14.0~$\le$~$K_\textnormal{s}$~$\le$~20.0~mag and $-$0.2~$\le$~$(Y - K_\textnormal{s})$~$\le$~1.0~mag; the normalised $\chi^2$ is used to characterize the goodness-of-fit, defined as
\begin{equation}
\chi^2=\dfrac{1}{m-k-1} \sum_i \dfrac{(n^i_{\rm model} - n^i_{\rm data})^2}{n^i_{\rm data}}
\end{equation}
where $m$ is the number of bins in the CMD, $k$~=~14 the number of free parameters to fit and $n^i_{\rm model}$ and $n^i_{\rm  data}$ are the number of stars in the $i^{\rm th}$ bin for the model and the observed CMD. We use a downhill-simplex algorithm to find the linear combination of SPMs that minimises the normalised $\chi^2$, plus the second safeguard as described in Section~4 of \citet{Harris2001} to avoid settling on a local rather than global minimum. By comparing the results for all pairs of $(m-M)_0$ and $A_V$, we finally obtain the best-fitting CMD with the smallest normalised $\chi^2$.

The double-distance model approach is similar, except that we use 14 SPMs to describe the nearer populations plus another 14 SPMs to describe the more distant ones. The distant  SPMs have the same ages, metallicities and extinctions as the nearer ones, but they have a distance modulus $(m-M)^{\rm far}_0$ =  $(m-M)^{\rm near}_0$ + 0.4 mag (0.4 mag is the average difference between the two RC peaks in the 4 sub-regions). All other parameters are the same as in the single-distance model.

The observed CMD for tile SMC~5\_6 and its best-fitting single- and double-distance models are shown in Fig.~\ref{d_fn.fig}. The best-fitting single-distance model is found for $(m-M)_0$~=~18.60~mag and $A_V$~=~0.25~mag and has a normalised $\chi^2$ of 2.19. The best-fitting double-distance model is found for $(m-M)^{\rm near}_0$~=~18.60~mag (thus $(m-M)^{\rm far}_0$~=~19.00~mag) and $A_V$~=~0.20~mag and has a normalised $\chi^2$ of 1.62. By comparing the normalised $\chi^2$, it is immediately apparent that the double-distance model fits the observation better than the single-distance one. Specifically, the double-distance model reproduces the elongated RC feature as seen in the observed CMD; in contrast, the single-distance model has a small and compact RC, which lies near the brighter component of the observed RC feature. The single distance model tries to produce the fainter RC component with a modest colour shift with respect to the brighter RC, which is not present in the observed CMD. Also the predicted magnitude difference between the two clumps is smaller than the observed one. Thus our modelling supports the idea that the double RCs arise from stellar populations located at two distances. 

Such a simple test is not adequate to show the effect of line-of-sight depth in the central regions, where the brighter clump is not distinct from the fainter clump. A future analysis based on VMC data including a more detailed description of the distance distributions would provide vital information about the geometry of the SMC.

\begin{figure}
\centering
 \hspace{-1.5cm}
 \includegraphics[height=0.38\textwidth,width=0.55\textwidth]{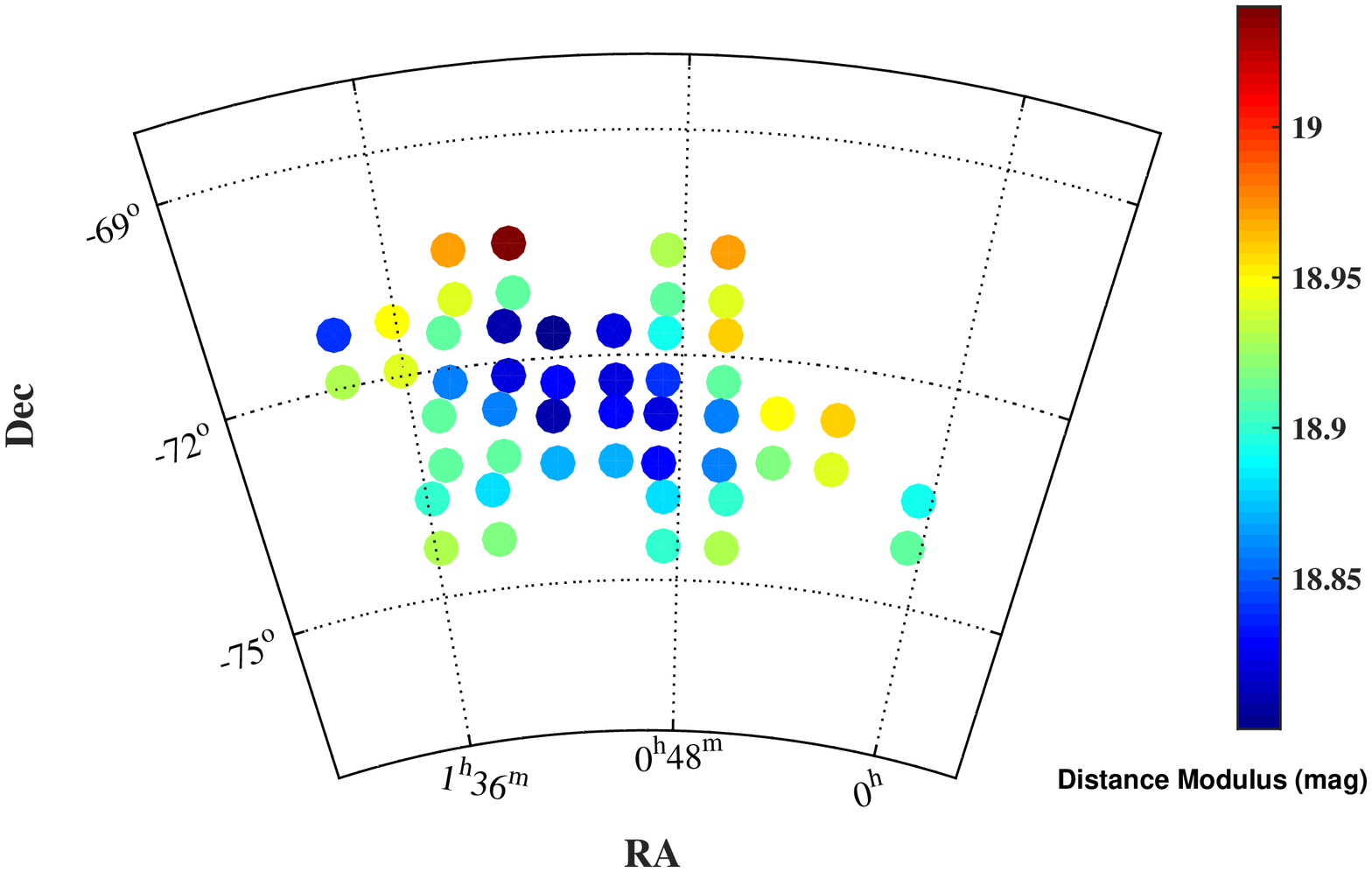} 
 \caption{Two-dimensional map of the mean distance modulus to the sub-regions in the SMC based on the fainter clump (the narrow component in the central regions and the single component in the south-western regions). }
 \label{meandm.fig}
\end{figure}

\begin{figure}
\centering
 \includegraphics[width=0.5\textwidth]{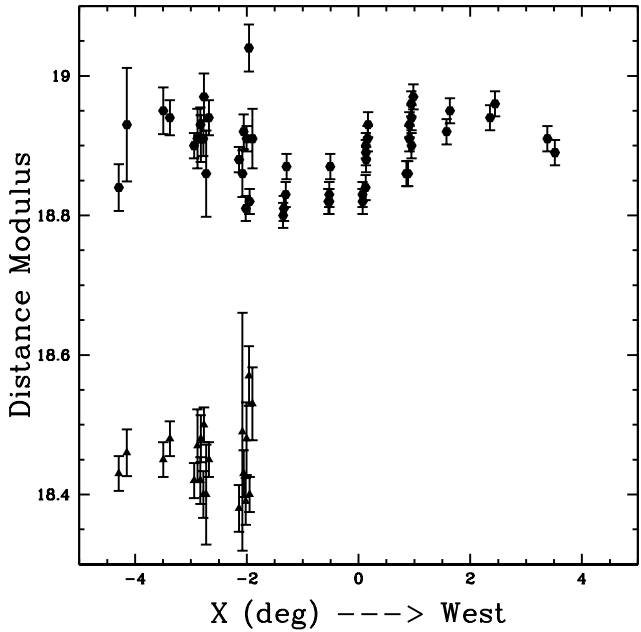} 
 \caption{Variation of the distance modulus in the east--west direction.}
 \label{eastwest}
\end{figure}

\begin{figure}
\centering
 \includegraphics[width=0.5\textwidth]{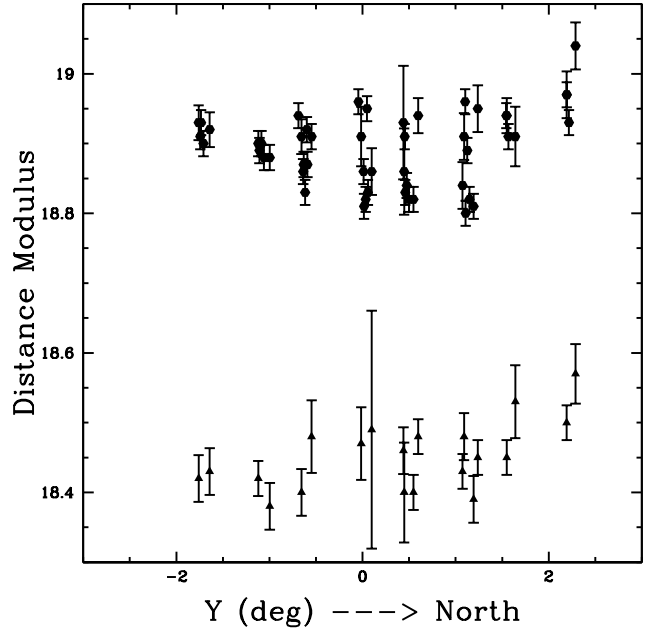}
\caption{Variation of the distance modulus in the north--south direction.}
\label{northsouth}
\end{figure}

\subsection{3D structure}

The extinction-corrected $K_\textnormal{s,0}$ magnitudes of the bright and faint clumps are converted to distance moduli using the absolute magnitude of the RC stars provided by \cite{Laney2012}. Using high-precision observations of solar neighbourhood RC stars, they provide the absolute RC magnitudes in the $J$,$H$ and $K$ bands in 2MASS system. We converted them into the VISTA $K_{S}$ system using the transformation relations provided by \cite{Rubele2015}. The absolute magnitude of RC stars in the VISTA $K_{S}$ band, $M_{Ks}$ = $-$1.604$\pm$0.015 mag. Based on this value, the extinction-corrected $K_\textnormal{s,0}$ magnitudes of the two components of the RC are converted to distance moduli. 

The absolute magnitudes of RC stars in the solar neighbourhood and in the SMC are expected to be different owing to the differences in metallicity, age and star formation rate between the two regions. This demands a correction for population effects while estimating the distance modulus to different sub-regions in the SMC. \cite{SG2002} estimated this correction term in the $K$ band to be $-0.07$ mag for the SMC. They simulated the RC population in the solar neighbourhood and in the SMC using stellar population models and including the star formation rate and age -- metallicity relation derived from observations. Then, they compared the difference in the absolute magnitudes in the two systems to quantify the population effects. We applied this correction ($-0.07$ mag) to all the 13 tiles in this study to estimate the distance moduli. 

The average distance modulus based on the peak magnitudes of the fainter clump (the narrow component in the central regions and the single component in the south-western tiles) is 18.89 $\pm$ 0.01 mag which is in agreement, within the uncertainties, with recent estimates of the distance to the SMC. The brighter clump stars in the eastern tiles are at an average distance modulus of $\sim$ 18.45$\pm$ 0.02 mag. The two-dimensional distance modulus map of the SMC based on the peak magnitudes of the fainter clump (the narrow component in the central regions and the single component in the south-western tiles) is shown in Fig.~\ref{meandm.fig}. The plot shows that the central regions are at a closer distance than the outer regions. 

The east--west and north--south variation of the distance moduli obtained from the peak magnitudes of the different components of the RC luminosity function are shown in Figs \ref{eastwest} and \ref{northsouth} respectively. The circles and triangles represent the distance moduli estimated from the peak magnitudes corresponding to the fainter clump (the narrow components in the central region) and the brighter clump in the eastern tiles, where they have a distinct peak. The black circles in Fig. \ref{eastwest} show a gradient from east to west, in the inner ($-2^\circ < X  < 2^\circ$) region. Such a gradient in the mean distance, with the eastern regions being at a closer distance, has been observed in previous studies of the inner regions of the SMC using RR Lyrae stars (\citealt{Haschke2012}; \citealt{SS2012}). The outer regions in the east (with distances based on the fainter clump) and the west are at similar distances. The eastern regions which show a distinct bright clump, are beyond $(X = -2^\circ)$ and they are closer to us than the SMC by $\sim$ 10 -- 12 kpc. As can be seen from Fig. \ref{northsouth}, there is no significant distance gradient in the north--south direction. 

We note that there could be a variation of population effects and the correction term may vary across the SMC. To quantify this variation, we generated the absolute RC luminosity functions for the two tiles SMC 4\_3 and SMC 5\_6, following the same procedure described by \cite{SG2002}. These two tiles are in the inner and outer regions of the SMC. For the tile, SMC 4\_3, we used the star formation history results from \cite{Rubele2015}. As illustrated in Sect.5.2, the results based on zero-depth model are not a good approximation for SMC 5\_6. So we used the results from the double distance model (Sect.5.2) for this tile. The difference between the magnitude corresponding to the peaks of the absolute RC luminosity functions of SMC 4\_3 and SMC 5\_6 is $\sim$ $-0.04$ mag. As this estimate is based on the peak of the absolute RC luminosity function, it represents the most abundant, classical  intermediate-age RC stars of the region. Our distance estimates are also based on the peak of the observed RC luminosity function. Hence the variation in the population correction term is less likely to be affected by the contribution of young RC stars. 

We do not expect this variation of $-0.04$ mag in population correction to affect our final results significantly, which are mainly based on the relative difference between the two RC peaks (in eastern tiles) in the same tile. Even some contribution to one of the clumps from the inner region would only increase the relative difference due to this variation in the population effects. The distance gradient observed in the inner regions may have some contributions from this variation. But a similar gradient based on RR Lyrae stars is observed in studies of the inner regions in the SMC. As indicated earlier, a detailed star formation history analysis including the effects of a distance spread is needed to accurately calculate the variation of the absolute magnitude of the RC across the SMC and we plan to address this in a future paper.

The presence of stars at a closer distance than the main body of the SMC in the eastern regions and a mild indication of distance gradient in the central regions could be due to tidal interactions resulting from the recent encounter of the MCs. This possibility is discussed in detail in the next section. 


\begin{figure}
\centering
 \includegraphics[width=0.5\textwidth]{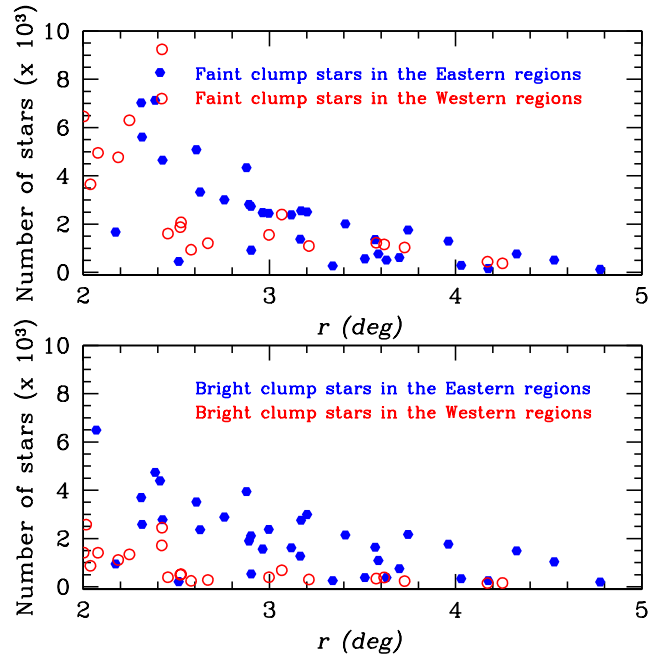}
\caption{The radial variations of the brighter (bottom-panel) and fainter (top-panel) RC stars in the outer ($\it{r}$ $>$ 2$^\circ$.0) eastern (RA $>$ 00h 52m 12s .5, shown in blue solid bullets) and western (RA $<$ 00h 52m 12s .5, shown in red open circles) regions are shown.}
\end{figure}

\begin{figure*}
\centering
\begin{tabular}{cc}
\setlength{\tabcolsep}{0.0002mm}
 \hspace{-1.1cm}
 \includegraphics[height=0.4\textwidth,width=0.52\textwidth]{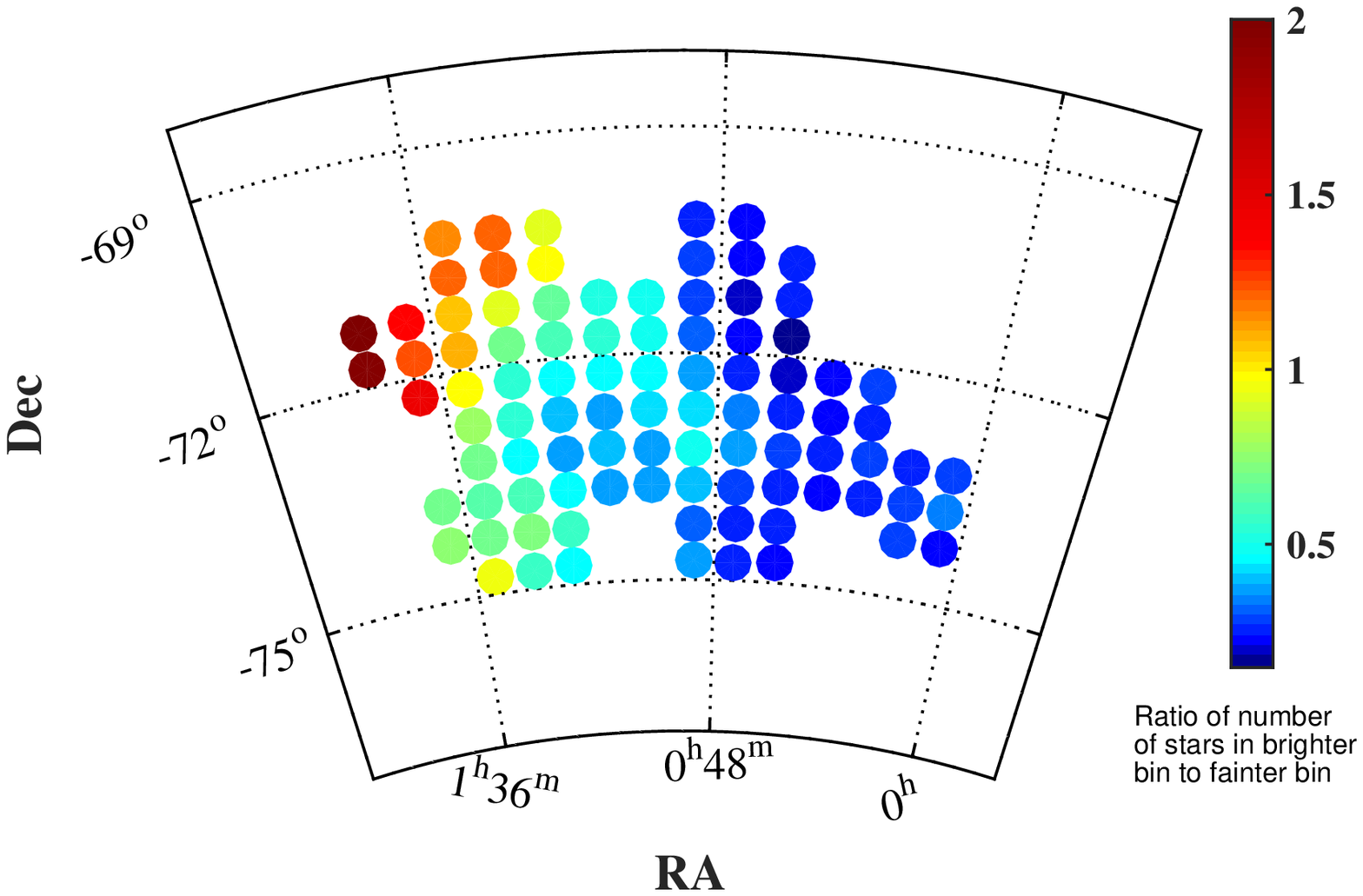} &
 \includegraphics[width=0.5\textwidth]{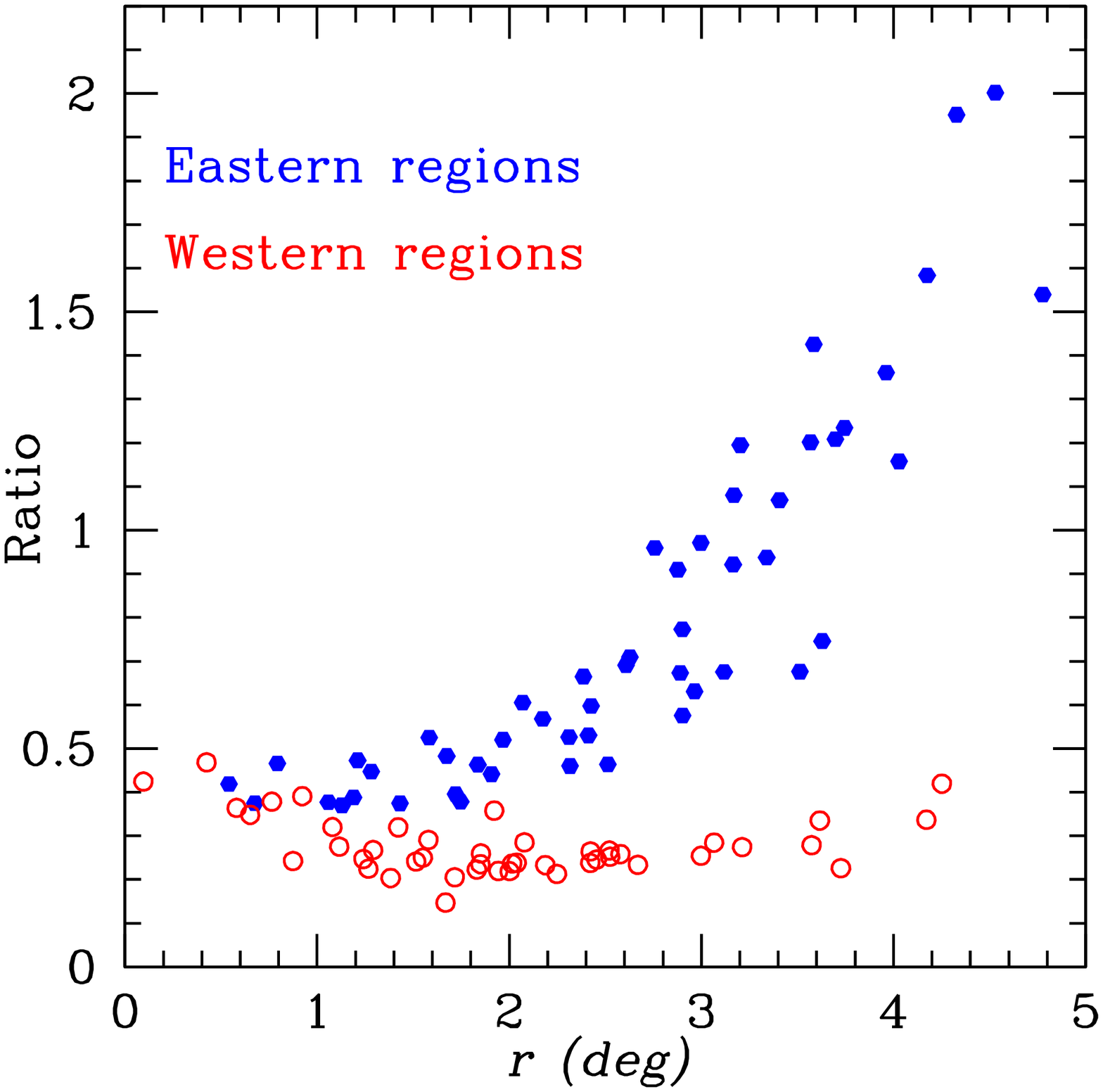}\\
 \end{tabular}
\caption{The left-hand and the right-hand panels show the spatial distribution of the ratio of bright RC stars (in the magnitude range, 16.8 $\le$ $K_\textnormal{s,0}$ $\le$ 17.1 mag) to the faint RC stars (in the magnitude range, 17.25 $\le$ $K_\textnormal{s,0}$ $\le$ 17.55 mag) and the radial variation of the fraction respectively.  In the right-hand panel, the blue solid bullets and the red open circles represent the eastern and the western regions, respectively.}
\end{figure*}

\begin{figure}
\includegraphics[width=0.5\textwidth]{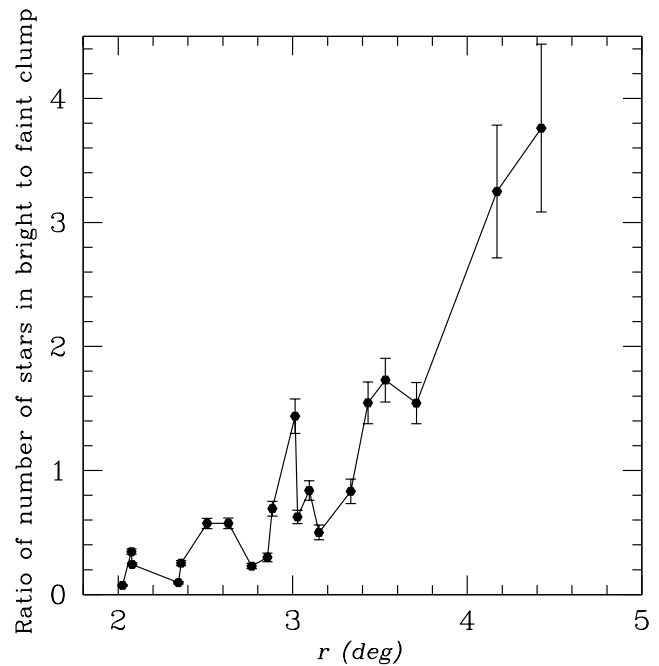}
\caption{Ratio of stars in the brighter clump to the fainter clump in the sub-regions of the eastern tiles (SMC 3\_5, SMC  4\_5, SMC 5\_5, SMC 6\_5 and SMC 5\_6 where the two clumps are distinct in the CMD as well as in the luminosity function) plotted against the radius corresponding to the centre of the sub-region.}
\end{figure}

\section{Effect of tidal interaction}

The present study suggests the presence of RC stars in front of the main body of the SMC in the galaxy's eastern regions. The comparison of the fraction of the bright clump stars in the eastern regions to the RC stars in other regions (in the same magnitude range of the bright clump) is essential to understand the nature of their origin. From Fig. 6 we can see that only in the eastern regions the two clumps are distinct. In all other regions, there is overlap in the luminosity functions of the fainter and brighter clumps. To understand the spatial variation of the two clumps, we divided the observed region into smaller bins of $\sim$ 0.4 $\times$ 0.5 deg$^2$ area. Then in these regions, we divided the RC spanning the colour range, 0.55 $\le$ $\it{(Y-Ks)_{0}}$ $\le$ 0.8 mag into the fainter (17.25 $\le$ $K_\textnormal{s,0}$ $\le$ 17.55 mag ) and brighter (16.8 $\le$ $K_\textnormal{s,0}$ $\le$ 17.1 mag) clumps. We note that the number of bright clump stars, especially in the central regions ($\it{r}$ $<$ 2$^\circ$.0), is an upper limit as as it could be influenced by young bright RC.

The radial variations of the brighter and fainter clump stars in the outer ($\it{r}$ $>$ 2$^\circ$.0) regions are shown in the two panels of Fig. 12. The top-panel shows that the fainter clump stars have similar distributions in the eastern and western regions. The bottom-panel shows that the brighter clump stars are more concentrated in the eastern regions than the western regions. The east--west asymmetric distribution of brighter clump stars rules out the possibility of them being the extended halo of the SMC. Although there are more brighter clump stars in the eastern regions, their number distribution decreases towards larger radii. If the foreground stars have their origin in the LMC, then we expect the number distribution of the brighter clump to decrease as the distance from the LMC increases. We see the reverse trend, which suggests that the foreground stars have their origin in the SMC.  Thus, the most viable explanation for the origin of this foreground population in the eastern and central regions of the SMC is tidal stripping of SMC stars.  This naturally explains the origin of the MB as caused by tidal stripping of material from the SMC. 

Tidal effects are expected to be stronger radially outwards from the centre of the SMC. The two-dimensional plot of the fraction of stars in the brighter (16.8 $\le$ $K_\textnormal{s,0}$ $\le$ 17.1 mag) magnitude bin to the fainter (17.25 $\le$ $K_\textnormal{s,0}$ $\le$ 17.55 mag) magnitude bin and the radial variation of the fractions in the eastern and western regions are shown in the left-hand and right-hand panels of Fig. 13. The two plots clearly show an increase of brighter clump stars towards larger radii in the outer ($\it{r}$ $>$ 2$^\circ$.0) eastern regions. The estimated fractions based on the number of stars in the brighter and fainter magnitude bins are approximate as we do not consider the actual profile of the RC distribution. 
Thus, the right-hand panel of Fig. 13 suggests a global trend of the variation of the fraction, but the numbers may not be accurate. 

A more accurate separation of the brighter and fainter clumps is possible in the eastern sub-regions shown in Fig. 6, where the two clumps are distinct and we have the corresponding profile fits. From the profile fits to the RC luminosity function, we can estimate the actual number of RC stars in the brighter and fainter clumps and also the total number of RC stars. We investigated the variation of the number of RC stars, in the sub-regions of the eastern tiles, as a function of radius. The ratio of stars in the bright clump to that in the faint clump (the fraction is estimated only for the sub-regions in the eastern fields which show two distinct peaks in the luminosity function) is plotted against radius in Fig. 14. The fraction of RC stars in the brighter clump gradually increases with radius and it becomes $\sim$ 3 -- 4 times the number of RC stars in the fainter clump at $\sim$ 4$^\circ$. 

The increase of the fraction of the brighter RC stars in the total clump with radius on the eastern side of the SMC indicates that the effect of tidal interaction is more significant in regions away from the centre of the SMC. The number ratio of the brighter to fainter clump in each sub-region corresponds to the mass ratio of the foreground populations to the main body. The ratio presented here from the inner to the outer regions in the eastern SMC traces the mass of the stripped population during the last encounter of the MCs and can be used as an observational constraint to dynamical models which try to understand the formation and interaction history of the Magellanic system.  

\citet{Noel2013, Noel2015} found that the properties (based on synthetic CMD techniques) of the intermediate-age stellar population in the MB are similar to those of the stars in the inner 2.5 kpc region of the SMC. However, they compared the properties of stars in the MB with fields in the south and west of the inner SMC. \cite{Noel2013} obtained  some quantitative estimates of the dynamical tidal radius at the pericentric passage using the formula given by Read et al. (2006),\\\\
$R_\textnormal{t} = [M_\textnormal{{SMC}}/M_\textnormal{{LMC}}]^{(1/3)} ({r_\textnormal{p}} - R_\textnormal{t})$\\\\
where ${r_\textnormal{p}}$ is the pericentric distance during the interaction. They assumed ${r_\textnormal{p}}$ = 5 kpc and a mass ratio of 1 and obtained a dynamical tidal radius of 2.5 kpc.  If we adopt the MCs mass ratio of 1:10 (which is more realistic based on recent studies by \cite{Besla2015} and $r_\textnormal{p}$ = 6.6 kpc \citep{DB2012}, then the tidal radius is $\sim$ 2.1 kpc. These values match the radius from which we start seeing the effect of the interaction in the form of two distinct RC luminosity functions in the eastern region of the SMC. 
Model 2 of \cite{Besla2012} considers a direct collision of the MCs during the last encounter and this model explains the observed structure and kinematics of the MCs better than Model 1 of \cite{Besla2012}, which considers only a close encounter of the MCs and not a direct collision. Model 2 predicts the removal of stars and gas from the deep potential of the SMC and formation of the MB during a direct collision of the MCs $\sim$ 100-300 Myr ago. In this model, where the MCs experience a direct collision, the pericentric distance is close to zero. Then   tidal interactions can affect the inner 2 kpc of the SMC. The presence of a broad component and a mild distance gradient in the central regions could be owing to this effect. However, because of the strong gravitational potential the stripping is not as efficient as in the outer regions. Hence, we do not see the bright clump feature as a distinct component in the central regions. Thus, tidal effects explain the observed bimodality in the RC luminosity function in the eastern regions ($\it{r}$ $\sim$ 2$^\circ$ -- 4$^\circ$) of the SMC.

\section{Discussion}
We found a tidally stripped intermediate-age stellar population in the eastern regions of the SMC, 2$^\circ$.5 to 4$^\circ$ from the centre,  $\sim$ 11.8 $\pm$ 2.0 kpc in front of the SMC's main body.  The central regions show large line-of-sight depth and a distance gradient towards the east. These observed features are most likely to be due to the tidal effects during the recent encounter of the MCs  100 -- 300 Myr ago. These results provide observational evidence of the formation of the MB from tidal stripping of stars from the SMC and the number ratio of the bright to faint RC stars can be used to constrain the mass of the tidally stripped component.  

 Some of the observed features could have a partial contribution from the variation of population effects of RC stars across the SMC. But our simple modelling of the observed CMD supports the idea that the double RC features in the eastern tiles arise from stellar populations located at two distances. A future analysis based on the the entire VMC data set, including the detailed distance distributions will disentangle the line-of-sight depth and population effects in the RC luminosity function.

Studies based on other distance tracers such as RR Lyrae stars (Muraveva et al., in prep.) and Cepheids (\citealt{JD2016}; \citealt{Ripepi2016}; Ripepi et al., in prep.) in similar regions of the SMC show only a distance gradient, with the eastern regions being closer to us than the western regions. These studies do not show a clear distance bimodality. On the other hand, young star clusters in the eastern regions of the SMC are found to be at a closer distance ($\sim$ 40 kpc; \citealt{Bica2015}). We see a distance gradient in the inner regions and a distinct RC population in the outer regions which are closer to us. The absence of bimodality in some of the distance tracers could be due to their low density in the outer regions. The RC stars are numerous and are distributed homogeneously across the SMC and, hence, show the bimodality clearly. 

These observed features in the present study can provide new constraints on theoretical models of the Magellanic system and can also be used to validate existing models. The most recent models of the Magellanic system based on new proper-motion estimates and incorporating a more realistic MW model are by \cite{DB2012} and \cite{Besla2012}. Both models explain most of the observed features of the Magellanic system but are meant to reproduce the gaseous features of the system. \cite{Nid2013} analysed these models and found that they could not reproduce the observed large line-of-sight depth and bimodality in their $\it{r}$ = 4$^\circ$ fields in the eastern regions of the SMC. We did not find evidence of a stellar population behind the SMC corresponding to the counter bridge in the eastern fields, as suggested by the disk model of \cite{DB2012}. The counter bridge is predicted to be at distance of $\ge$ 75 kpc which is $\sim$ 15 kpc behind the main body. The detection of a fainter peak ($Ks_0$ $\sim$ 17.8 mag) in the RC luminosity function corresponding to the RC population of the counter bridge is difficult because of the contamination by RGB stars, especially if the density of the RC population in the counter bridge is lower. 

\cite{SS2015} found a few extra-planar Cepheids, behind the disk, in the eastern region of the SMC.  This suggests the presence of young stars in the counter-bridge region. The non-detection of RC stars in the counter-bridge suggests that the counter-bridge may be devoid of intermediate-age populations. The pure spheroidal model of the SMC by \cite{DB2012}, which represents more intermediate-age stellar features, does not prominently show the presence of a counter-bridge. The two models in the simulations of \cite{Besla2012} appear to show the presence of a counter-bridge. In Model 1 the bridge and counter-bridge regions are populated with similar stellar densities, whereas in Model 2 the counter-bridge is less populated than the bridge region. Thus, the non-detection of the intermediate-age RC stellar population corresponding to the counter-bridge and the presence of tidally stripped stars from the central regions of the SMC support a direct collision of the MCs during the last encounter.

A detailed chemical and kinematic study of the RC stars in the SMC is needed to provide more constrains on the tidal stripping model. Recently, \cite{Parisi2016} analysed the radial velocities and the metallicities of RGB stars in the SMC clusters and field. They found a negative metallicity gradient in the inner regions ($\it{r}$ $< $ 4$^\circ$) and then an inversion in the outer regions ($\it {r}$ $\sim$ 4$^\circ$ -- 5$^\circ$) in the direction of the MB. One of the possibilities for this inversion in metallicity gradient would be the presence of metal-rich stars in the outer regions which are tidally stripped from the inner regions of the SMC. Compared with the RGB stars, the RC stars provide a handle on their distances and are hence more appropriate to verify this phenomenon. 

\section{Summary and Conclusions}
We present a study of the intermediate-age RC stars in the inner 20 deg$^2$ region of the SMC using VMC survey data in the context of the formation of the MB. The mean distance modulus to the main body of the SMC is 18.89 $\pm$ 0.01 mag. A foreground population (11.8 $\pm$ 2.0 kpc) from the inner ($\it{r}$ $\sim$ 2$^\circ$) to the outer ($\it{r}$ $\sim$ 4$^\circ$) regions in the eastern SMC is identified. The most likely explanation for the origin of the foreground stars is tidal stripping from the SMC during the most recent encounter with the LMC. The radius ($\it{r}$ $\sim$ 2 -- 2.5 kpc) at which the signatures become evident/detectable in the form of distinct RC features matches the tidal radius at the pericentric passage of the SMC. These results provide observational evidence of the formation of the MB from tidal stripping of stars from the SMC and the number ratio of the bright to faint RC stars can be used to constrain the mass of the tidally stripped component.  A detailed chemical and kinematic study of the RC stars would provide better constraints to the tidal stripping scenario. \\\\

{\bf Acknowledgements}: We thank the Cambridge Astronomy Survey Unit (CASU) and the Wide Field Astronomy Survey Unit (WFAU) in Edinburgh for providing calibrated data products under the support of the Science and Technology Facility Council (STFC) in the UK. S.S acknowledges research funding support from Chinese Postdoctoral Science Foundation (grant number 2016M590013). S.R was supported by the ERC Consolidator Grant funding scheme (project STARKEY, G.A. n. 615604). R.d.G is grateful for research support from the National Science Foundation of China through grants, grants 11373010, 11633005 and U1631102. M.R.C acknowledges support from the UK's Science and Technology Facilities Council (grant number ST/M00108/1) and from the German Academic Exchange Service (DAAD). This project has received funding from the European Research Council (ERC) under the European Union's Horizon 2020 research and innovation programme (grant agreement No 682115). Finally, it is our pleasure to thank the referee for the constructive suggestions.

\end{document}